\documentclass[10pt,twocolumn,5p]{elsarticle}

\usepackage{hyperref}
\usepackage{balance}
\usepackage{enumitem}
\usepackage{listings}
\usepackage{booktabs}
\usepackage{array}
\usepackage{xcolor}
\definecolor{lstkeyword}{rgb}{0.10,0.10,0.60}
\definecolor{lstcomment}{rgb}{0.00,0.45,0.10}
\definecolor{lststring}{rgb}{0.62,0.10,0.10}
\lstdefinelanguage{yaml}{
  sensitive=false,
  comment=[l]{\#},
  commentstyle=\color{lststring}\itshape,
  morestring=[b]',
  morestring=[b]",
  stringstyle=\color{lstcomment},
  moredelim=**[il][\color{lstkeyword}\bfseries]{:},
  literate={:}{{\color{black}:}}1
           {---}{{\color{lstkeyword}-{-}-}}3
           {|}{{\color{lstkeyword}|}}1,
}
\usepackage{pgfplots}
\pgfplotsset{compat=1.18}
\usepgfplotslibrary{dateplot}
\usepackage{tikz}
\usetikzlibrary{positioning}
\usetikzlibrary{shapes}
\usetikzlibrary{fit}

\journal{arXiv}

\begin{document}

\begin{frontmatter}
  \title{Modern Software Engineering in the LAMMPS Molecular Dynamics
    Software Package: Development Process, Refactoring, Testing, and Deployment}

  \author[1]{Axel Kohlmeyer\corref{cor1}}
  \ead{axel.kohlmeyer@temple.edu}
  \cortext[cor1]{Corresponding author}
  \affiliation[1]{organization={Institute for Computational Molecular Science, Temple University},
    addressline={Science Education and Research Center (035-07)}, 
    city={Philadelphia},
    postcode={19122}, 
    state={PA},
    country={USA}}

  \begin{abstract}
    We review changes made in recent years to the development process of
    the LAMMPS simulation software package and the software itself.  We
    discuss how those changes have impacted the effort and workflow
    required to develop and maintain a software package that has been in
    existence for more than 30 years and where a significant part of the
    code base is contributed by external developers.  We also look into
    how those changes in the process have affected the code quality and
    ease of modifying and extending the software while at the same time
    its audience has changed from a cohort with a strong software
    development background to a group including many researchers with
    limited software development skills.  We explore how this
    contributes to LAMMPS' continuing growth in popularity over that
    period.  Specific topics include the source-code management and
    contribution workflow, automated testing and static analysis,
    security and supply-chain integrity, the
    refactoring and modernization of the code base (including the
    adoption of modern C++), the C, Python, and Fortran library
    interfaces, the build system and deployment, the documentation, the
    LAMMPS-GUI graphical interface, and our recent experiences using AI
    tools.  We
    highlight changes of the most recent development cycles and close
    with an outlook on future steps.
  \end{abstract}

  \begin{keyword}
    molecular dynamics\sep software engineering\sep high performance
    computing\sep testing\sep best practices\sep modern C++\sep build
    system\sep software packaging\sep software security\sep supply-chain
    integrity\sep collaborative software development
  \end{keyword}
\end{frontmatter}


\section{Introduction and Overview}
\label{sec:intro}

LAMMPS\cite{lammps_home} stands for \textsl{\textbf{L}}arge-scale
\textsl{\textbf{A}}tomic/\textsl{\textbf{M}}olecular
\textsl{\textbf{M}}assively \textsl{\textbf{P}}arallel
\textsl{\textbf{S}}imulator and is a classical molecular dynamics (MD)
simulation code with a focus on materials modeling.  It has many
features that are not easily available in other MD simulation software
packages to perform a wide variety of particle-based simulations at
different time and length scales.

LAMMPS is designed to run efficiently on a wide variety of parallel
computers using domain decomposition\cite{lammps95} and MPI message
passing\cite{mpi_forum}, and to be easy to extend and modify.  In
addition to MPI parallelization throughout, it supports
multi-threading\cite{openmp}, vectorization, and GPU acceleration for
varying parts of its functionality\cite{lammps22}.

LAMMPS was originally developed at Sandia National Laboratories, a
facility of the US Department of Energy (DOE), but currently distributed
LAMMPS releases\cite{lammps_releases} include contributions from
\emph{many} research groups and individuals from institutions around the
world.  Most of the funding for maintaining LAMMPS has come and is still
coming from the US Department of Energy.

\begin{figure}[bt]
  \begin{tikzpicture}
  \pgfplotsset{every tick label/.append style={font=\scriptsize}}
  \begin{axis}[
      width=0.98\linewidth,
      height=34ex,
      ybar stacked,
      bar width=6pt,
      xmin = 2003.3,
      xmax = 2027.3,
      ymin = -100,
      ymax = 1200000,
      xtick align=inside,
      xtick ={2005, 2008, 2011, 2014, 2017, 2020, 2023, 2026},
      x tick label style={/pgf/number format/.cd,%
          scaled x ticks = false,
          set thousands separator={},
          fixed},
      ytick={50000, 250000, 500000, 750000, 1000000},
      y tick label style={/pgf/number format/.cd,%
          scaled y ticks = false,
          set thousands separator={,},
          fixed},
      ymajorgrids,
      ylabel={\textsf{Lines of Code}},
      ylabel style={font=\scriptsize,xshift=-3ex,yshift=-2.5ex},
      legend pos=north west,
      legend style={font=\scriptsize,xshift=4ex,yshift=-0.7ex},
    ]
    \addlegendimage{empty legend}
    \addplot[draw=black,fill=yellow!60!gray] table [x=Year,y={Fortran},col sep=comma] {figure1-data.csv};
    \addplot[draw=black,fill=green!50!gray] table [x=Year,y={CUDA},col sep=comma] {figure1-data.csv};
    \addplot[draw=black,fill=red!70!gray]   table [x=Year,y={Python},col sep=comma] {figure1-data.csv};
    \addplot[draw=black,fill=blue!60!gray]  table [x=Year,y={C++},col sep=comma] {figure1-data.csv};
    \legend{\textsf{\textbf{Language:}},\textsf{Fortran},\textsf{CUDA},\textsf{Python},\textsf{C/C++}};
    \node[rotate=90, fill=white] at (axis cs:2026.8,540000) {\textsf{\tiny (Data as of Summer 2026)}};
    \node[fill=white,rotate=90] at (axis cs:2004.05,280000) {\textsf{\scriptsize LAMMPS 99}};
    \node[fill=white,rotate=90] at (axis cs:2005.05,320000) {\textsf{\scriptsize LAMMPS 2001}};
  \end{axis}
\end{tikzpicture}

  \caption{Lines of code in Fortran, CUDA, Python, and C/C++
  in the LAMMPS package over time (counted with cloc\cite{cloc_github})}
\label{fig:git_lines}
\end{figure}
Over a period of more than 30 years, LAMMPS has significantly evolved,
same as computer hardware, operating systems, programming languages, and
build systems have changed with it.  The growth of the package over the
years, measured by counting the lines of source code in the entire
package\footnote{These counts do not include the lines of code in
  external libraries required by packages like ML-QUIP(libAtoms),
  KIM(kim-api), PLUMED(Plumed2), or ML-PACE(ACElib) to name a few} is
shown in figure~\ref{fig:git_lines}.  With the added source code come
additional features and contributions\cite{lammps22} from developers
within and outside the core group.  This broadens the range of research
topics that can be addressed and has been recognized with an R\&D~100
award\cite{rnd100_2018} in 2018.

\begin{figure*}[htb]
\input{figure2}
\caption{Citations\cite{google_scholar} of the LAMMPS overview
  articles\cite{lammps95,lammps22} over time connected with milestones in
  LAMMPS' history (Sect.~\ref{sec:history})}
\label{fig:citations}
\end{figure*}

In addition to the original publication\cite{lammps95} describing the
message passing parallelization used in LAMMPS, a second LAMMPS overview
article\cite{lammps22} was published in 2022 that summarizes many of the
added features and improvements since LAMMPS was converted from Fortran
to C++.  The number of annual citations for both publications that
describe the core functionality of LAMMPS shown in
figure~\ref{fig:citations} illustrates how LAMMPS has not only grown in
size but also grown in popularity over time.

This growth of LAMMPS has at times exposed limitations of the
development process and thus prompted adjustments.  A major milestone in
this regard was the move to use the git\cite{git} distributed source
code management system and to have the project publicly hosted on
GitHub\cite{lammps_github}.  Through this step, the maintenance burden
for the software is shared between contributors and other core LAMMPS
developers.  Another significant milestone was connecting the GitHub
project with automated tests, so that many issues with contributions in
pull requests are detected. Those issues \emph{must} be resolved to
unblock the pull requests from being merged into the distribution.

Also, for any such large software project, there is always a conflict
between conservative developers and progressive developers, and
maintainers have to find a balance between those two approaches for the
software to remain useful and relevant.  This requires regular exchange
between the core developers and written-down policies and protocols for
the project management, so that knowledge can be easily shared with
junior developers and that there is a point of reference for choices
that have been made in case there are disagreements.

In this paper\footnote{An earlier and much shorter version of this work
  was presented at the 2025 US-RSE conference~\cite{lammps_usrse25}.},
we first review the history and evolution of the LAMMPS project and its
source code (Sect.~\ref{sec:history}).  We then discuss the development
process and source-code management (Sect.~\ref{sec:scm}), the automated
testing and static-analysis infrastructure (Sect.~\ref{sec:test}), the
security and supply-chain practices (Sect.~\ref{sec:security}), the
refactoring and modernization of the code base
(Sect.~\ref{sec:refactor}), the C, Python, and Fortran library
interfaces (Sect.~\ref{sec:library}), the build system and deployment
(Sect.~\ref{sec:build}), the improvements to the documentation
(Sect.~\ref{sec:docs}), our experience using AI in LAMMPS development
(Sect.~\ref{sec:ai-usage}), and the LAMMPS-GUI application
(Sect.~\ref{sec:gui}).  Throughout, we point out the lessons learned,
and we close with a summary (Sect.~\ref{sec:summary}) and an outlook on
planned future developments (Sect.~\ref{sec:future}).

\section{History of LAMMPS}
\label{sec:history}

\begin{figure*}[ht]
\begin{minipage}[c]{0.75\linewidth}
\begin{tikzpicture}
    \pgfplotsset{every tick label/.append style={font=\scriptsize}}
    \begin{axis}[
      width=\linewidth,
      height=35ex,
      date coordinates in=x,
      ybar,
      bar width=3pt,
      ymajorgrids,
      legend pos=north west,
      legend style={font=\scriptsize},
      y tick label style={/pgf/number format/.cd,%
          scaled y ticks = false,
          set thousands separator={},
          fixed},
      xticklabel=\year,
      xtick={2006-01-01, 2008-01-01, 2010-01-01, 2012-01-01, 2014-01-01, 2016-01-01, 2018-01-01, 2020-01-01, 2022-01-01, 2024-01-01, 2026-01-01},
      xtick align=inside,
      xmin = 2006-10-01,
      xmax = 2026-12-01,
      ymax = 1550,
      legend image code/.code={\draw[#1, draw=black] (0cm,-0.1cm) rectangle (0.2cm,0.1cm);}
    ]
    \addplot[draw=black,fill=blue!60!gray]  table [x=Quarter,y={Commits},col sep=comma] {figure3-data.csv};
    \legend{\textsf{Commits per quarter (excluding merge commits)}}
    \node[rotate=90] at (axis cs:2026-09-15,950) {\textsf{\tiny (Data as of Summer 2026)}};
\end{axis}
\end{tikzpicture}
\end{minipage}
\begin{minipage}[c]{0.25\linewidth}
\textsf{
\scriptsize
Git repository and Github statistics:\\[-0.5ex]
\centering{\tiny (approximate, as of the 2026 stable release)} \\[1ex]
\begin{tabular}[b]{lr}
  Total source lines: & 2,000,000 \\
  Total lines of code: & 1,430,000 \\
  Total files: & 9,350 \\
  Total source files: & 6,950 \\
  Optional packages: & 94 \\
  Total git commits: & 50,300 \\
  Contributors: & 500 \\
  Total pull requests: & 3,765 \\
  Issues submitted: & 1,275 \\
  GitHub Stars: & 2900 \\
  Forks on GitHub: & 2000
\end{tabular}
}
\end{minipage}

\caption{Graph showing git commits per quarter to the main LAMMPS
  branch\cite{lammps_github} (left) and some source code related
  statistics (right)}
\label{fig:git_commits}
\end{figure*}

To better understand some of the quirks and design decisions in the
LAMMPS source code, it is helpful to review the history of the LAMMPS
project and the source code.  Figure~\ref{fig:citations} correlates the
LAMMPS timeline with milestones in LAMMPS' history, most of which are
discussed in the following paragraphs.  Starting at about 2010, we can
see a steady increase in the number of annual citations, which indicates
a continuing growth of the LAMMPS user base, which has continued through
2025.  This growth provides motivation to the LAMMPS core developers to
continue on the chosen path to keep LAMMPS an accessible, reliable, easy
to learn, and easy to modify software.

\subsection{Origins Using Fortran}
\label{sec:history:fortran}

The development of LAMMPS began in the mid-1990s under a cooperative
research \& development agreement (CRADA) between two DOE laboratories
(Sandia and Lawrence Livermore) and three companies (Cray, {Bristol
  Myers Squibb}, and DuPont).  The goal was to develop a large-scale
parallel classical MD code; the coding effort was led by Steve Plimpton
at Sandia and a first LAMMPS publication\cite{lammps95} described the
core parallelization approach using domain decomposition. After the
CRADA ended, a final version called LAMMPS 99 was released, which was
written in Fortran 77.  This version was available under a no-cost
license, but was not open source.

As LAMMPS development continued at Sandia, its memory management was
converted to Fortran 90 and new versions were released; the final LAMMPS
version written in Fortran 90 was released as LAMMPS 2001.  The final
versions of both Fortran variants are still available for download on
the LAMMPS homepage\cite{lammps_download} as historical references.
However, for the remainder of this publication, we focus on the C++
versions that followed.

\subsection{Conversion to C++ and Modern C++}
\label{sec:history:cplusplus}

The current LAMMPS software is a rewrite in C++ and was first publicly
released as an open source code in 2004.  It includes many new features
beyond those in LAMMPS 99 or 2001.  The main motivation for using C++
was to take advantage of polymorphism (which was not available in
Fortran at the time) and C-language pre-processing to largely automate
the process of including optional features into a LAMMPS executable,
and thus to allow building custom LAMMPS executables that include only
features required by the individual researcher.  Due to limitations in
portability and reliability of advanced C++ features on the platforms
supported by LAMMPS, especially advanced parallel platforms available
at that time, the programming style of LAMMPS at this time was best
described as ``C with Classes''.

Over the years, many new capabilities were added to LAMMPS, while
maintaining core functionality and code design. Deviations were only
accepted in optional features called ``packages''.  The new capabilities
include new models and interfaces to external packages, but also support
for accelerators like GPUs, multi-threading, and vectorization.  Using
``packages'' for contributed features allowed the developers to keep the core of the
LAMMPS code relatively lightweight despite the significant growth of the
overall source code in the distribution as shown in figure~\ref{fig:git_lines}.
As of the 2026 LAMMPS stable release, the distribution consists of
approximately 750 core source files with almost 160,000 lines of code
and 94 optional ``packages'' with an additional 6,200+ source files
containing approximately 1,250,000 lines of code for a total of about
1,430,000 lines of code\footnote{Line counts according to cloc
  v2.08\cite{cloc_github}}.  Figure~\ref{fig:git_commits} shows some
additional general statistics for the LAMMPS source code and GitHub
repository in the list on the right.

A notable addition is the integration of features from the GRASP MD
program developed by Aidan Thompson, which started in 2006.  This made a
variety of many-body potentials available for the first time in an
open-source software package.  The next substantial addition was the GPU
package in 2009, an effort that combined contributions from multiple
groups and was led by W. Michael Brown\cite{gpu2011,gpu2012}.  This
marked the first availability of GPU acceleration in LAMMPS (using CUDA
or OpenCL, later extended to HIP). Another significant milestone was the
addition of the KOKKOS package\cite{lammps_kokkos} in 2014.  It was
originally written by Christian Trott and provides
\emph{performance-portable} support for accelerator hardware on all
major HPC platforms by employing the Kokkos
library\cite{kokkos,kokkos22} (Fig.~\ref{fig:citations}) and has grown
since then significantly thanks to a variety of contributors.

The development was made more transparent by providing access to a
read-only mirror of the subversion\cite{subversion} repository used for
LAMMPS development which was also converted to a read-only git
repository\cite{git}.  The maintenance of LAMMPS during this time was
shared between researchers at Sandia and Temple University.  After many
discussions, the LAMMPS development moved in 2016 to GitHub, and the
LAMMPS developers use git directly and exclusively.  GitHub would
maintain a subversion client gateway until January 2024.  The mechanics
of this transition are described in Sect.~\ref{sec:scm:git}.

Subsequently, the moratorium on using features from the C++ standard
library was partially lifted.  The C-language library ``stdio'' is still
strongly preferred over C++ ``iostreams'' for reading and writing files.
Instead, the \{fmt\} formatting library\cite{libfmt} (in anticipation of
the \texttt{std::format} functionality included in the C++ 20 standard)
has been bundled for flexible and extensible formatting (see
Sect.~\ref{sec:refactor:fmt}).  This started a significant refactoring
effort in order to modernize and simplify the code base, which is
ongoing.  The refactoring was done not only for C++ but also for the
C-language library interface and the Python and Fortran modules that use
this interface, and finally was also extended to preparing the included
documentation in HTML and PDF format.  Since the ``19 March 2020''
version, LAMMPS requires a compiler that is at least compatible with the
C++11 standard, and since the ``10 September 2025'' version, LAMMPS
requires the C++17 standard.  We decided to skip C++14 because the
differences to C++11 are relatively small.

\subsection{Expansion of Core Development Team}
\label{sec:history:expand}

As for most large scientific software packages, maintaining the software
is a big challenge since there is very little credit to be had for
maintenance compared to adding features and publishing results with new
science that these new features enable.  Thus, the LAMMPS development
team is actively trying to distribute the load across more shoulders,
and by 2026 the core development team has grown to 12 developers from
five different institutions in the US and Europe.

In order to coordinate the development effort, we perform monthly video
calls (duration $\approx 1$ hour) for the core developers where
different aspects of the LAMMPS project are discussed.  In addition,
longer in-person meetings, often including prominent external LAMMPS
contributors or high-profile users, are occasionally organized.
Typically, these are planned to coincide with in-person LAMMPS Users
workshops in Albuquerque, or the annual meeting of the LAMMPS developers
from Temple University with the core developers at Sandia in New Mexico.

Finally, we have created an invitation-only Slack for LAMMPS developers,
which provides more direct access to individual LAMMPS developers and
provides monitoring of activities on GitHub and the automated testing
servers.  This is kept deliberately invitation-only to avoid its abuse
by (inexperienced) LAMMPS users. For those and others, there exists a
dedicated forum on the \href{https://matsci.org/lammps/}{MatSci.org}
Discourse server, which is the successor to the \emph{lammps-users}
mailing list that was hosted on SourceForge from 2005 until 2022.

The move to a larger and more diverse development team is also reflected
in the move of the project from a website hosted at Sandia to the
\texttt{lammps.org} domain, which is currently hosted by the Institute
of Computational Molecular Science (ICMS) at Temple University.  This
provides more flexibility and fewer restrictions on who can have access,
what information can be provided, what services can be run, and how
those services can be connected to other web services like GitHub or
Slack. Currently, the \texttt{lammps.org} domain hosts a web server for
the LAMMPS homepage, the LAMMPS documentation, documentation for
pre-compiled Windows installers, download servers, and the LAMMPS
testing front end.  Through the \texttt{lammps.org} domain, multiple
\texttt{@lammps.org} email addresses are also provided that are
forwarded to one or more LAMMPS core developers.

\section{Source Code Management}
\label{sec:scm}

From the very beginning, LAMMPS development has followed a continuous
delivery paradigm: that is, any released LAMMPS version \emph{and} the
public `develop' branch (Fig.~\ref{fig:git_branches}) are expected
to be always fully functional, and in case any issues arise, they are
addressed as quickly as possible to recover the full functionality.

\subsection{Custom Scripts and Subversion}
\label{sec:scm:svn}

In the early days, a new LAMMPS version was started by copying the
source tree.  Then changes were applied, and when a new version was
complete, custom scripts were run to create and upload a patch file with
the changes and a compressed tarball of the sources plus the manual in
HTML and PDF formats.  Concurrent development of features would require
multiple copies and integrating patches as they were released.  This is
not very different from the current \emph{distributed} development
process, only that it occurred exclusively on the computer of the lead
developer.  External developers would have their own copies (downloaded
from the LAMMPS homepage) and then send in new and modified files.

At this time, new patches were released frequently, often on subsequent
days or only a few days apart.  In effect, each patch represents an
added or updated feature or a fixed bug.  With the growing importance of
LAMMPS (inside and outside of Sandia), a subversion\cite{subversion}
repository for LAMMPS was started on a protected server in a Sandia data
center.  This would function as both an archive for the development
history and a backup of the sources, but also as the canonical point of
reference while providing concurrent write access to multiple developers
at Sandia.

\subsection{Move to Git and GitHub}
\label{sec:scm:git}

A practical problem of having a subversion server hosted by Sandia is
that it is not accessible to LAMMPS contributors outside of Sandia.
Thus, a process was established to create a read-only mirror of the main
LAMMPS development branch hosted on a publicly accessible server at
Temple University.  This server also converted the subversion repository
to a git\cite{git} repository, which accommodated the at the time
quickly growing number of developers who prefer git over subversion.

Since these were read-only mirrors, any changes would still have to be
emailed to a developer at Sandia to be added and committed to the
original subversion server.  Those would then be mirrored and show up on
the public subversion and git mirror servers within 24 hours.  However,
with frequent patch releases, this still sometimes led to a situation
where new patches would create new issues before bug-fix patches were
processed.  Additionally, contributed features would not always be
included immediately, depending on the availability of a LAMMPS
developer with write access to the subversion repository, and thus they
could be outdated already when released to the public.

To make the development process more transparent and to better
accommodate the needs of outside developers, the canonical LAMMPS
repository was eventually moved to use git and
GitHub\cite{lammps_github}.  Instead of emailing changes to a LAMMPS
developer for inclusion, changes are now submitted as pull requests on
GitHub.  Crucial for the development process is that pull requests can
be easily and quickly updated and tested (Sect.~\ref{sec:test}).
Potential merge conflicts but also test failures are detected, flagged,
and need to be resolved before merging becomes possible.

\subsection{Releases and Branches}
\label{sec:scm:releases}

\begin{figure}[tb]
  \resizebox{\columnwidth}{!}{
\begin{tikzpicture}
\tikzstyle{branch}=[
    rounded rectangle,
    fill=white,
    draw=black,
    line width=2pt,
    inner sep=8pt,
]
\tikzstyle{tag}=[
    rectangle,
    fill=white,
    draw=black,
]
\node[rectangle,draw,align=center,line width=2pt] (prs) {\textsf{Pull Requests}\\\textsf{on GitHub}};
\node[branch,below=of prs] (develop) {\textsf{`develop' branch}};
\node[branch,right=30ex of develop] (maintenance) {\textsf{`maintenance' branch}};

\node[branch,below=12ex of develop] (release) {\textsf{`release' branch}};
\node[branch,below=12ex of maintenance] (stable) {\textsf{`stable' branch}};

\node[tag,below=12ex of release,line width=1pt] (patch_tag) {\texttt{patch\_<date>}};
\node[tag,right=of patch_tag,line width=1pt] (stable_tag) {\texttt{stable\_<date>}};
\node[tag,right=of stable_tag,line width=1pt] (update_tag) {\texttt{stable\_<date>\_update<num>}};

\draw[->,>=stealth,line width=2pt] (prs) -- node[left] {\textsf{Merge commits}} (develop);
\draw[->,>=stealth,line width=2pt] (develop) -- node[above] {\textsf{Backports of bugfixes}} (maintenance);
\draw[->,>=stealth,line width=2pt] (develop) -- node[left,align=center] {\textsf{Feature release}\\\textsf{(every 4-8 weeks)}} (release);
\draw[->,>=stealth,line width=2pt] (release) -- node[above,align=center] {\textsf{Stable release}\\\textsf{(once per year)}} (stable);

\draw[->,>=stealth,dashed, line width=2pt] (release) -- node[left,yshift=0.25cm] {\textsf{Feature release}} (patch_tag);
\draw[->,>=stealth,dashed, line width=2pt] (stable) -- node[left,yshift=0.25cm] {\textsf{Stable release}} (stable_tag);
\draw[->,>=stealth,dashed, line width=2pt](stable) -- node[right,yshift=0.25cm] {\textsf{Update release}} (update_tag);

\draw[->,>=stealth,line width=2pt] (maintenance) to[bend left] node[right,align=center] {\textsf{Updates to}\\\textsf{stable release}} (stable);
\draw[->,>=stealth,line width=2pt] (stable) to[bend left] node[left,align=center] {\textsf{Reset on}\\\textsf{stable release}} (maintenance);

\node[draw,rectangle,dotted,inner sep=8pt,label={[align=center]below:{\textsf{Release tags}}},fit=(patch_tag) (stable_tag) (update_tag)] (reltags) {};

\end{tikzpicture}
}

  \caption{Flowchart showing relationships between the main branches in
  the LAMMPS git repository\cite{lammps_github} and conventions for applying
  release tags}
\label{fig:git_branches}
\end{figure}

As mentioned earlier, LAMMPS patches used to be released frequently,
which caused problems for people packaging LAMMPS or installing it on
clusters or supercomputers, since it would be impossible to provide
packages for each of those releases, and there was no indication whether
a specific release had particularly few known issues.

In the 2010s, the LAMMPS developers released on the order of 80--100
patch versions per year.  This high frequency was in part a consequence
of the development model of the time: changes were committed and
released directly, so that a patch which introduced a new problem often
had to be followed quickly by one or more bug-fix patches.  Starting in
2014, 2--5 of these patch versions per year would be retroactively
labeled as `stable' about a month after their respective release if no
known issues had emerged.  After the move to GitHub in 2016
(Sect.~\ref{sec:scm:git}), formal releases became progressively less
frequent---from about 30 per year in 2017 to roughly 5 per year today.
The `develop' branch---into which pull requests are merged---now plays
the role that frequent patch releases did in the early days, but changes
reach it both less often and more reliably, because each contribution
is tested and revised \emph{before} it is merged (Sect.~\ref{sec:test})
rather than corrected after release.  Developers now simply need to
follow that branch to stay up to date with development.  Instead,
regular releases are now called \emph{feature releases} and are done
only every 4--8 weeks when a sufficient number of new features and
improvements have accumulated.  These releases are tracked by the
`release' branch.

Starting in 2021, \emph{stable releases} are made only once per year
after a 1--2 month period in which only bug fixes are applied.  These
stable releases are tracked on the `stable' branch.  In addition, most
bug fixes to the `develop' branch are also back-ported and added to a
`maintenance' branch.  This `maintenance' branch is occasionally merged
with `stable' for \emph{stable update releases}; these continue until
the next stable release is made. Fig.~\ref{fig:git_branches} illustrates
the relationship between the four branches.  Stable releases and updates
are now the primary targets for the various options for LAMMPS
packaging.  This still maintains the spirit of the original LAMMPS
development effort, aiming to have all releases fully functional and
differ only in available features.

Currently, there are no long-term support (LTS) versions due to the
additional workforce requirements.  Stable releases receive updates only
until the next stable release.  With the significant changes introduced
in LAMMPS version 10 September 2025, we are considering maintaining the
22 July 2025 stable version for longer than a year to support users
running on environments that are not supported by version 10 September
2025 and later LAMMPS versions.  Most likely, only fixes for
\emph{significant} bugs will be back-ported to such an LTS version.

\subsection{Maintenance Policies and Conventions}
\label{sec:scm:maintenance}

As shown in Fig.~\ref{fig:git_commits} the number of commits has risen
from about 100--200 per quarter after the creation of the subversion
repository in 2006, to about 300--500 per quarter after the public
subversion (and git) mirror was made available in 2010, and then
increased sharply during the COVID-19 pandemic to about 600--1200
commits per quarter, a high level that has been sustained since.  This
commit history is a measure of development activity of LAMMPS.  Only the
move to git as source code management tool and using the GitHub pull
request management have allowed us to scale the development process to
this level of activity.  Now, contributions are improved and corrected
(if needed) \emph{before} they are included in the development branch
and the review process distributes the work across the entire LAMMPS
development team.  In addition, we have given recurring contributors
permission to triage incoming new issues and pull requests to further
engage the most competent person to provide a review.  This step is
largely automated by maintaining a \texttt{.github/CODEOWNERS} file that
maps the GitHub accounts of LAMMPS contributors to files or folders in
the LAMMPS source code (typically files they contributed) and
automatically requests reviews from those users.  Using AI based code
review using GitHub Copilot (Sect.~\ref{sec:ai-usage:review}) helps
to further lighten the load on the LAMMPS core developers.

\begin{table*}[tb]
\centering
\small
\setlength{\extrarowheight}{0.5ex}
\begin{tabular}{@{}>{\raggedright\arraybackslash}p{0.23\textwidth}p{0.46\textwidth}>{\raggedright\arraybackslash}p{0.24\textwidth}@{}}
\toprule
Section and test category & What it checks & When it runs \\
\midrule
\ref{sec:test:integration}: Integration testing & Compilation of LAMMPS across many operating systems, compilers, build settings, and package combinations. & On each pull request update \\
\ref{sec:test:unittest}: Built-in ``unit'' tests & Unit tests of utility functions and classes, test classes computing, changing, or consuming forces for regressions, validate granular classes, test C, Python, and Fortran interfaces. & On each pull request update, additional tests after merge \\
\ref{sec:test:regression}: Example regression tests & Runs complete example input decks and compares their thermodynamic output against reference logs. & After merge or on request \\
\ref{sec:test:static}: Static code analysis & \emph{Coverity Scan}, \emph{CodeQL}, Compiler warnings, \texttt{ClangSA}, and \texttt{clang-tidy} flag defects and questionable code constructs. & weekly, after merge, or nightly \\
\ref{sec:test:style}: Code-quality and style & Sanitizers, \emph{Valgrind}, \emph{IWYU}, \texttt{clang-format}, and custom scripts to enforce style conventions. & On demand and when enabled in build configuration \\
\bottomrule
\end{tabular}
\caption{Overview of the kinds of automated tests and code-analysis tools used for LAMMPS, what each verifies, and when it runs (Sect.~\ref{sec:test}).}
\label{tab:testing}
\end{table*}

To maintain consistency, we use branch protection for the `develop'
and `stable' branches so that \emph{all} changes must be submitted
as pull requests.  Accidental commits to a local copy of the `develop'
branch will be refused when pushing the branch.  This ensures that
all changes go through the automated review and testing process
(discussed in more detail in Sect.~\ref{sec:test}).

We also noticed that it is preferable to have a designated core
developer performing merges and handling releases.  That person may be
different for different releases.  Individual LAMMPS developers
``chaperone'' specific pull requests, and signal this by assigning the
pull request to themselves.  If they assign the pull request to the
developer responsible for merging, it is the signal that the pull
request is ready to be merged (from the perspective of the new or
improved functionality that is added) and can be subjected to a final
round of testing and enforcing various formal and programming style
requirements.

Furthermore, an agreement has been reached that at least \emph{two} core
LAMMPS developers have checked a pull request before it is merged.
Typically, that would be the developer performing the merge (which
counts as an implicit approval) and one more developer.  This is in part
enforced by the project configuration that requires at least one
approval from a core developer with write access to the repository and
in part by the convention that the developer designated to perform
merges must not be the only approver of a pull request.  Also, if a
developer wants to (temporarily) block merging a specific pull request,
a request for changes with a suitable comment can be made.  This is
regularly done by core developers that have particular expertise or
responsibility for a subsystem within LAMMPS so they can delay the
merging of a pull request until they have had sufficient time to review
the changes.  This convention was specifically established to
accommodate the situation that none of the core developers currently can
work on LAMMPS full time and thus may have limited availability at
times.

It is also an important policy of the core development team that any
significant changes must be made in consensus.  To facilitate such
discussions as well as discussions about other organizational issues
like planning of upcoming releases, workshops, or development sprints,
monthly meetings are held using videoconferencing.  These meetings are
also an opportunity to transfer knowledge about the maintenance process
from senior to junior members of the development team as we try to grow
the core development team to prepare for retiring senior LAMMPS
developers.

\section{Automated Testing}
\label{sec:test}

As mentioned in Sect.~\ref{sec:scm:git}, the move to git and GitHub
already helped improve code quality by introducing code reviews
\emph{before} merging.  Another step that has been \emph{extremely}
effective in preventing bugs and compilation issues from being
introduced into the code base is that changes in the GitHub hosted
LAMMPS repository and submitted pull requests trigger a variety of
automated actions, specifically several kinds of tests for different
LAMMPS configurations and build environments.  Most of these tests are
run when pull requests are submitted or updated, thus indicating
problems to the contributor to resolve.  Failed tests prevent pull
requests from being merged until \emph{all} issues are resolved and all
tests pass.  The tests are currently mostly run in the Microsoft Azure
cloud using GitHub workflows and additional tests on servers at Temple
University.  In this section, we discuss the different kinds of test
runs and what their benefits are (Table~\ref{tab:testing}).

\subsection{Integration Testing}
\label{sec:test:integration}

LAMMPS is designed to be compatible with different CPU hardware, many
operating systems (mainly Linux variants, Windows, or macOS, but also
FreeBSD, Solaris, AIX), using several compilers (for example, GNU,
Clang, Intel, Microsoft VC++, Nvidia, IBM), two different build systems,
and with a variety of build settings (with a static or shared library,
or with 32-bit or 64-bit atom IDs and image flags).  The main
restrictions in practice are currently the requirement for fully C++17
standard compliant compiler toolchain and having 64-bit integer data
types.  Also, there are dependencies between some optional packages and
classes where changes to a base class can break compilation of the
derived class in a package.

A recurring problem with contributed code in pull requests is that the
contribution compiles and runs correctly in the development environment
of the contributor with the settings used by that developer, but can
fail in one of the other supported environments.  Possible causes are
incorrect use of some code constructs, lack of testing during
development, writing of non-portable code, or limitations of or
unavailability of external libraries on all platforms and so on.

To ensure that a pull request does not break compilation of LAMMPS in
general, but for all existing features in particular, a wide variety of
compilations are triggered after a pull request submission or any
update.  These run currently in the Microsoft Azure cloud using GitHub
runners on Linux, Windows and macOS.

\subsection{Built-in Test Library}
\label{sec:test:unittest}

Testing whether LAMMPS compiles at all is only one prerequisite to
ensure that LAMMPS behaves as expected on all supported platforms.  To
test its functionality the \texttt{unittest} tree was added to LAMMPS in
2020 and populated with a variety of tests.  Some of those tests are
unit tests in the strict sense, and test utility functions and classes
that are used for convenience or portability and for recurring tasks.

A large number of the included tests are more like a hybrid between unit
and regression tests.  This is necessary since many features of LAMMPS
cannot be executed until a viable simulation system that uses physically
meaningful settings has been created.  Often, this system has to
have specific properties enabled or components allocated and populated
with reasonable data like per-atom charges or topology information
to be suitable for testing features in LAMMPS.

LAMMPS uses the GoogleTest C++ library as a test framework for C, C++,
and Fortran language code, where testing Fortran requires writing
suitable Fortran-to-C wrappers.  Testing is not enabled by default and
is only available by setting \texttt{-D ENABLE\_TESTING=ON} when
compiling LAMMPS using the CMake build system (more on that in
Sect.~\ref{sec:build:cmake}).  This includes the \texttt{unittest} tree
in the build process and also downloads and compiles a specific version
of GoogleTest\cite{googletest} that is compatible with LAMMPS' build
requirements.  After compilation is complete, tests can be run with the
CTest utility that is part of the CMake software package.  Tests for
individual features are configured such that they are skipped for LAMMPS
compilations that lack the corresponding optional packages or features.

Specifically, most of the test programs in the folders
\texttt{unittest/force-styles}, \texttt{unittest/granular}, or
\texttt{unittest/bpm} are tools that generate a variety of inputs
on-the-fly from fragments and settings provided in YAML files
(Listing~\ref{lst:yaml}) and compare the energies, stresses and forces
for an initial configuration and after a few MD steps against previously
recorded reference values.  If supported, the results of the
\texttt{single()} function are also compared to what the
\texttt{compute()} function produces.  Similarly, the writing and
reading of binary restart files and (text-mode) data files are tested,
and the resulting energies are compared.  Finally, different simulation
settings that follow different code paths, floating-point precision
settings, and communication patterns but should produce the same
energies and forces within a given epsilon are tested, and---if
available---accelerated variants of the same style (for the OPENMP, OPT,
KOKKOS (Serial, OpenMP, CUDA, HIP, SYCL), INTEL, or GPU package) are
tested as well, if the hardware supports it.

\begin{lstlisting}[language=yaml,float=tb,caption={Beginning of a
 \texttt{force-styles} test file (here for the \texttt{lj/cut} pair
 style; the reference data are abridged).  The YAML header sets the
 relative tolerance \texttt{epsilon}, the styles that must be available
 (\texttt{prerequisites}), the commands and coefficients that build
 the model, and the internal parameters to verify (\texttt{extract}).
 The recorded per-atom forces, energies, and stresses that the test
 program regenerates and compares against follow below.},label={lst:yaml}]
---
lammps_version: 22 Dec 2022
date_generated: Thu Dec 22 09:53:54 2022
epsilon: 5e-14
skip_tests:
tags:
prerequisites: ! |
  atom full
  pair lj/cut
pre_commands: ! ""
post_commands: ! |
  pair_modify mix arithmetic
  pair_modify shift yes
input_file: in.fourmol
pair_style: lj/cut 8.0
pair_coeff: ! |
  1 1  0.02   2.5
  2 2  0.005  1.0
  2 4  0.005  0.5
  3 3  0.02   3.2
  4 4  0.015  3.1
  5 5  0.015  3.1
extract: ! |
  epsilon 2
  sigma 2
natoms: 29
init_vdwl: 749.2470096189502
init_coul: 0
# ... reference stress and per-atom forces omitted ...
\end{lstlisting}

Those \texttt{force-styles} tests are intentionally designed to be very
sensitive. Properties such as atomic forces, global energies, and global
stresses must not have relative errors greater than a global
\emph{epsilon} parameter (typically defined in the range of
$1\cdot 10^{-13}$ to $1\cdot 10^{-8}$) when compared to reference data
included in the test's YAML file. The \emph{epsilon} parameter can be
adjusted in the YAML file for individual tests to accommodate pair
styles with a larger amount of ``noise'', e.g. due to using
interpolation tables.  The global \emph{epsilon} parameter is modified
by empirical factors for different test settings, including accelerator
packages and lower precision force kernels where a correspondingly
larger error is expected.  The same reference data are used for all test
variants, which has led to the detection and correction of a significant
number of bugs and incomplete or inconsistent implementations.  This
comprehensive coverage of all pair-style variants, including their
accelerated versions, was also what made the large-scale, AI-assisted
refactoring of the embedded atom method (EAM) pair styles
(Sect.~\ref{sec:ai-usage:refactor}) safe to carry out: because every
format and accelerator combination is checked against the same reference
data, any deviation introduced by the consolidation would have been
caught immediately.  More generally, the thorough testing required to
validate such a refactoring will sometimes also expose pre-existing,
unrelated bugs.

Still, some tests are more sensitive than others and may fail on
different platforms or when compiling LAMMPS at a high optimization level.
Other tests may take an excessive time under some conditions.
The YAML files for the tests include a keyword for setting tags like
``unstable'' or ``slow'' that may be used to tell the CTest utility via
command-line flags to selectively skip such tests as needed.

In the same spirit as the \texttt{force-styles} tests, we recently added
a comprehensive collection of unit and validation tests for discrete
element modeling (DEM) simulations (see Sect.~\ref{sec:ai-usage:develop}
item~\ref{sec:ai-usage:develop:grantest}).  The important difference
from the previous category of tests is that these tests not only detect
regressions from a recorded reference state, but also compare simple
benchmark test cases against the corresponding analytical solutions.
This kind of validation is of high importance in the DEM simulation
community and thus adding them addresses a significant deficiency of
LAMMPS for those applications and complements the significant
development that the DEM support in LAMMPS has seen in recent years.

Another set of tests in the \texttt{unittest} tree concerns the C-language
library interface and its wrappers for Python and Fortran.  These tests
primarily focus on features that are specific to the interfaces and thus
not covered by the tests for force styles and similar.

Finally, there are also custom tests for individual LAMMPS commands or
to check the format of files generated with the dump command or of files
that are parsed as potential files or molecule files.  Where possible,
tests are performed not only to get expected results with \emph{correct}
input but also to confirm that LAMMPS aborts with the expected error message
when given incorrect input (also known as ``death tests'').

Due to the size of the LAMMPS code base, the test coverage is
incomplete.  At the time of this writing about 80\% of the files and
classes are covered and about 45\% of the lines of source code.  Most of
the untested code, though, is in packages and features that are not used
frequently, so that for many common use cases of LAMMPS the test
coverage is much better than the percentages indicate.  The biggest
deficit is for compute styles and fix styles that are not related to
force computation or time stepping.  This is mostly because of the
additional programming effort required since it is not effective to use
a generic test tool like the force-style tests do.

\subsection{Regression Testing Using Examples}
\label{sec:test:regression}

In regression testing, the result of a given input deck is compared with
reference results for the same input created with an older version of
the software which are assumed to be correct.  Although not originally
intended for that purpose, the LAMMPS \texttt{examples} tree provides a
large pool of resources for that purpose, since examples typically
include complete input decks and output logs to run those inputs with 1
and 4 MPI tasks.  There are several challenges here: a) the submitted
examples may be too large or use too many MD simulation steps and
therefore may take a long time to complete; b) the examples do not
contain a sufficient amount of so-called thermodynamic output; c) the
impact of the demonstrated feature is not reflected in the thermodynamic
output; d) the simulation contains randomized elements that are not
reproducible across platforms; and e) the example will diverge because
of exponentially growing accumulated noise from the limitations of
floating-point math\cite{floating_point}.

With the intention to not ``pollute'' the \texttt{examples} tree with
the modifications needed for regression tests, a separate tree with
copies of the examples was used where inputs are adjusted as needed and
examples not suitable for regression testing are removed or only tested
for completion of the run without checking the output.  This approach
provided us with a large library of regression tests quickly, but it
proved to be a significant effort to maintain it, considering the pace
at which LAMMPS development advances.  As a result, the regression test
set of examples has diverged from the examples in the LAMMPS source
distribution and is missing most recent contributions.  Most recently
the hardware at Temple running these tests had to be retired and now
only variants of those tests specifically aimed at testing styles in the
KOKKOS package remains and these tests are run internally at Sandia
without reporting their status back to GitHub.

Thus, a new effort is underway to use more advanced scripts for
regression testing that can use the examples from the LAMMPS source
distribution directly and require only some non-invasive modifications
to a few examples to make them compatible.  This has the twofold benefit
that new tests for new features are automatically discovered when
corresponding examples with suitable reference log files are added, and
that the existing example outputs will be kept consistent with the
current LAMMPS version.

Testing the complete collection of example inputs for regressions is
still time consuming, and thus only performed \emph{after} the merge
of a pull request or after adding a specific label to the pull request.
The new regression test tool in development speeds up the processing by
distributing the tests across multiple runners.  It also has a ``quick
mode'' suitable for testing pull requests; that mode first checks which
commands and styles are changed by the pull request, and then only runs
regression tests on examples containing those commands or styles.  In
``quick mode'' also the maximum number of tests is curbed; if the list
is larger than a given threshold, only a randomly chosen subset is
tested.

\subsection{Static Code Analysis}
\label{sec:test:static}

Part of the testing strategy to eliminate common bugs and improve code
quality for LAMMPS is the use of multiple tools using static code
analysis.  The first such step is the option to compile LAMMPS with
verbose warnings enabled (e.g., using \texttt{-Wall -Wextra} with the
GCC or Clang compilers).  During the parsing process of the compiler,
often deeper insights into the code structure may be obtained, and
certain kinds of logic errors or typos can be identified.  For that to
work well, all easily avoidable warnings have to be eliminated (e.g.,
about unused variables) so that any remaining warnings stand out more.
It is also important to use different compilers, since those tend to
detect different issues differently well.

\begin{figure}[tb]
\begin{tikzpicture}
    \pgfplotsset{every tick label/.append style={font=\scriptsize}}
    \begin{axis}[width=\columnwidth,height=30ex,
      legend pos=north west,
      legend style={font=\tiny},
      ymajorgrids,
      y tick label style={/pgf/number format/.cd,%
          scaled y ticks = false,
          set thousands separator={},
          fixed},
      date coordinates in=x,
      xticklabel=\year,
      xtick={2021-01-01, 2022-01-01, 2023-01-01, 2024-01-01, 2025-01-01, 2026-01-01},
      xmin = 2020-09-01,
      xmax = 2026-03-01,
      ymin = -100,
      ymax = 3200,
      ]
      \addplot[sharp plot,draw=blue!60!gray,thick]  table [x=Date,y={Fixed Defects},col sep=comma] {figure5-data.csv};
      \addplot[sharp plot,draw=red!60!gray,thick]  table [x=Date,y={Outstanding Defects},col sep=comma] {figure5-data.csv};
      \legend{\textsf{Fixed Defects},\textsf{Outstanding Defects}}
      \node[rotate=90] at (axis cs:2026-01-15,1750) {\tiny\textsf{(Data as of December 2025})};
\end{axis}
\end{tikzpicture}

\caption{Graph of defects reported by Coverity Scan\cite{coverity}}
\label{fig:scan_defects}
\end{figure}

A more straightforward approach to static code analysis is provided by
tools such as \emph{Coverity Scan}\cite{coverity} or
\emph{CodeQL}\cite{codeql}.  These perform a more thorough pattern-based
and heuristic analysis of the source code, and thus take significant
time to process and analyze it.  For that reason, the \emph{CodeQL}
analysis is performed only after merging, and the (most time-consuming)
processing with \emph{Coverity Scan} only once per week (unless
explicitly requested).  Given the programming conventions used in
LAMMPS, many of which predate the conversion from Fortran to C++, there
is a significant number of false positives, especially about class
members that are not initialized in the constructor\footnote{In this
  specific case we found adding initializers for scalars to be not
  always helpful as they could hide whether those members are properly
  set to their physically meaningful values before use in the later
  steps of setting up a simulation system and the missing initialization
  would not be flagged by tools like Valgrind\cite{valgrind}.}

Using \emph{Coverity Scan} in particular has been immensely useful in
detecting hard-to-find issues such as data type conversion errors,
copy-and-paste errors, logic errors, mismatched \texttt{new/delete/malloc()/free()},
dead code, and much more. Fig.~\ref{fig:scan_defects} lists the number of detected and
resolved problems in the last five years as identified by \emph{Coverity
  Scan}.  Static code analysis is applied not only to C++ code but also
to Python code. Using the \emph{CodeQL} tool has been especially
beneficial in this regard.

Recently, we added another static code analysis check using tools
from the Clang/LLVM project through the CodeChecker Python package\cite{codechecker}.
This package runs selected checks from the \texttt{clang-tidy} command and the Clang
static code analyzer and has prompted additional corrections. Through \texttt{clang-tidy},
this also includes suggestions for modernizing C++ code constructs (for example,
to use \texttt{auto} to avoid redundant and potentially inconsistent type information).

We also experimented early on with AI-based code review through the GitHub
Copilot tool, with initially underwhelming results: while a few reports were
helpful, many suggestions were not useful or were incorrect, and often there
were none at all.  With better project-specific instructions this has since
improved considerably, and AI-assisted review is now part of our regular
workflow, as discussed in Sect.~\ref{sec:ai-usage}.

\subsection{Tools for Code Quality \& Programming Style}
\label{sec:test:style}

The CMake build system for LAMMPS has provisions to enable a number of
additional tools to analyze and improve the LAMMPS source code.  For
example, there is support for \emph{clang-tidy} and
\emph{include-what-you-use} (IWYU), which are based on the LLVM compiler
infrastructure\cite{llvm,llvm_web} and are capable of detecting
incorrect or outdated C++ code constructs and inconsistent use of
include statements, respectively.

The GCC and Clang compilers both also support the creation of instrumented
executables using the \texttt{-fsanitize} flag.  A selection of sanitizers
are available, for instance, for undefined behavior, memory leaks,
or invalid memory access.  These are less detailed and accurate than
the Valgrind memory checker, but execution is much faster.

We also provide custom suppression files for the Valgrind memory checker
tool, so that (ideally) only errors within LAMMPS are flagged.  Especially
the MPI and OpenMP runtime libraries trigger a large number of false
positives, and the provided suppressions hide most of them.  Valgrind
has been particularly effective in detecting memory management issues
when used in combination with built-in unit testing via CTest.  For
this purpose, the so-called ``death tests'', which would also lead to
many false positives, can be excluded from the test runs.  A similar
option to use Valgrind\cite{valgrind} or instrumented executables is
available for the regression checking tool that is currently in development.

Finally, there are some custom scripts that check source files if
they conform to various documented conventions.  The same applies to
configuration files for the clang-format tool.  Both promote a homogeneous
and consistent format and programming style.
These conventions are explained in the ``Programmer's Guide'' part of
the LAMMPS manual (Sect.~\ref{sec:docs:developer}).


\section{Security, Integrity, and Supply-Chain Practices}
\label{sec:security}

LAMMPS is a user-level scientific application rather than a networked
service: it runs simulations from local input and links against a range
of external---and sometimes experimental---research libraries.  It
therefore reads largely untrusted input and performs file I/O on the
user's behalf, and, for performance, omits validation in its inner loops,
so that malformed or malicious input can crash it or trigger unintended
file operations.  For this reason LAMMPS must never be run with elevated
privileges, and most reports that initially look like a security
vulnerability turn out to be ordinary bugs.  These expectations, together
with the channel for reporting suspected vulnerabilities, are stated in a
documented security policy (the \texttt{SECURITY.md} file in the
repository).  In practice, ``security'' for a package like LAMMPS rests
on three pillars: the correctness and memory safety of the code, the
integrity and provenance of the sources and of the released binaries, and
the trustworthiness of a distributed development process to which many
external contributors submit changes.

Several mechanisms described earlier serve the first pillar directly.
Every change to the protected \texttt{develop} and \texttt{stable}
branches must arrive as a pull request and be reviewed by at least two
core developers (Sect.~\ref{sec:scm:maintenance}); the extensive
automated testing (Sect.~\ref{sec:test}) and the static analysis with
\emph{Coverity Scan}, \emph{CodeQL}, and \texttt{clang-tidy}
(Sect.~\ref{sec:test:static})---where \emph{CodeQL} runs its
\texttt{security-and-quality} query set---together with the memory
sanitizers and \emph{Valgrind} (Sect.~\ref{sec:test:style}) and the
AI-assisted review (Sect.~\ref{sec:ai-usage:review}) catch memory-safety
defects and questionable constructs, the dominant source of
vulnerabilities in C/C++ software, \emph{before} they reach a release.

The second and third pillars concern the chain of custody of the code.
Because all changes flow through reviewed pull requests on protected
branches, no single contributor can alter a release branch unilaterally,
and the \texttt{.github/CODEOWNERS} mapping
(Sect.~\ref{sec:scm:maintenance}) ties review of each file to identified
maintainers; release tags are cryptographically (GPG) signed.  To block a
class of source-level supply-chain attacks in which malicious code is
concealed using Unicode homoglyphs or bidirectional-override characters
(so-called ``Trojan Source'' attacks~\cite{trojan_source}, CVE-2021-42574), every contribution
is checked and only all-ASCII source code is accepted.

LAMMPS also actively manages its dependency supply chain.  Components that
are critical to the build are \emph{vendored}---a verified copy is bundled
with the sources at a pinned version---among them the \{fmt\} formatting
library, a JSON library, the Kokkos performance-portability library, a
fallback reference BLAS/LAPACK implementation, and the KISS FFT library.
External packages that are not bundled but can be downloaded automatically
by the build system (Sect.~\ref{sec:build:cmake}) are verified against
recorded SHA-256 checksums; in addition, the LAMMPS download server keeps
validated mirror copies of these dependencies and uses them as a fallback
when an upstream source is unavailable, so that a build never depends on
the continued availability or integrity of a third-party server.  An
automated dependency monitor keeps the pinned versions used by the
continuous-integration workflows up to date.

Finally, the integrity of what users actually download is protected.
Every file served from the \texttt{lammps.org} servers is accompanied by
a SHA-256 checksum, and since the ``10 September 2025'' release the LAMMPS
releases on GitHub are \emph{immutable}: once published, neither the
release tag nor the attached assets---the source tarball and the
pre-compiled LAMMPS and LAMMPS-GUI packages---can be altered, and GitHub
generates a build-provenance attestation that lets users verify each
artifact.  These technical measures are backed by the written governance
and maintenance policies (Sect.~\ref{sec:scm:maintenance} and
Sect.~\ref{sec:docs:policies}) and by the planned move to the High
Performance Software Foundation (Sect.~\ref{sec:future}), which together
are intended to keep this security posture sustainable as the project and
its contributor base continue to grow.

LAMMPS is itself an upstream dependency for many other software
ecosystems, so these integrity guarantees matter well beyond the project.
Its sources are repackaged by several Linux distributions and by the
source-based deployment and package managers that are widely used in
research computing, among them Spack\cite{spack},
EasyBuild\cite{easybuild}, Homebrew\cite{homebrew}, and the conda-forge
channel of the Anaconda/conda ecosystem\cite{condaforge}.  Such consumers
generally package the annual \emph{stable} releases together with their
periodic bug-fix updates rather than the more frequent feature releases
(Sect.~\ref{sec:scm:releases})---one of the purposes of that two-tiered
release model---and they have supply-chain-integrity requirements of their
own: stable, verifiable source archives with known checksums and a clear
chain of provenance.  The signed, immutable releases and the published
SHA-256 checksums described above are designed to satisfy exactly these
needs, so that the integrity of LAMMPS carries through to the many systems
built on top of it.

\section{Code Refactoring and Related Changes}
\label{sec:refactor}

\begin{table*}[tb]
\centering
\small
\setlength{\extrarowheight}{0.1ex}
\begin{tabular}{@{}p{0.27\textwidth}p{0.66\textwidth}@{}}
\toprule
Removed Packages or Tools & Replacement or reason for removal \\
\midrule
\texttt{REAX} (Fortran) & Superseded by \texttt{REAXFF}, an independent C/C++ reimplementation. \\
\texttt{MEAM} (Fortran) & Reimplemented in C++ with extensions, function by function (now the \texttt{MEAM} package). \\
\texttt{MESSAGE}, \texttt{LATTE} & Replaced by the more general \texttt{MDI} coupling package. \\
\texttt{USER-CUDA} & Evolved into the multi-architecture \texttt{KOKKOS} package. \\
\texttt{MPIIO}, \texttt{MSCG} & Removed as unmaintained or obsolete. \\
\texttt{ATC}, \texttt{AWPMD}, \texttt{POEMS} & Removed as unmaintained (a visible drop in the Coverity defect count, Fig.~\ref{fig:scan_defects}). \\
\midrule
\texttt{restart2data} & Replaced by a built-in command-line conversion.\\
\texttt{amber2lammps.py} & Obsolete (Python2 based); Replaced by external AMBER2LAMMPS package.\\
\texttt{xmovie} & Superseded by \texttt{dump image} and external visualization tools (VMD, OVITO). \\
LAMMPS-SHELL & Superseded by LAMMPS-GUI. \\
LAMMPS-GUI, moltemplate, i-PI & Unbundled; now maintained as separate projects. \\
\bottomrule
\end{tabular}
\caption{LAMMPS packages and bundled tools that have been removed or
  replaced over time (Sect.~\ref{sec:refactor:remove}).}
\label{tab:removed}
\end{table*}

While adoption of git (Sect.~\ref{sec:scm:git}) and extensive automated
testing (Sect.~\ref{sec:test}) are primarily aimed at making the
process of integrating new contributions into LAMMPS more efficient
without breaking existing functionality, there are also changes to the
existing LAMMPS code to refactor it (i.e. rewrite without changing
functionality) to modernize it and make it easier to maintain.  This
applies to different components of the LAMMPS source code and in this
section we are trying to categorize them. The refactoring process goes
hand in hand with unit testing (Sect.~\ref{sec:test:unittest}), as
it is advisable to first write tests to establish the status quo to
which any refactoring changes are compared.  Ideally, any refactoring
changes will reproduce the same behavior as before.

As mentioned in Sect.~\ref{sec:history:cplusplus}, the LAMMPS sources
are organized into basic functionality and optional ``packages''.
Originally, these packages were grouped into ``core'' packages (assuming that
they were well maintained by the LAMMPS core developers), ``user''
packages (maintained by external contributors), and a ``USER-MISC''
package for all kinds of one-off contributions.  Over time, the quality
of the code and the level of maintenance for different packages diverged
substantially.  After observing that the LAMMPS core developers were
often blamed for issues with ``user'' packages (which were supposed to
be maintained externally) and after some of them had been orphaned, a
concerted effort was started to eliminate this distinction.  Many of
the changes were motivated by reports of defects from static code analysis
(Sect.~\ref{sec:test:static}).  This effort culminated in a
reorganization and renaming of packages in Summer 2021: a) all ``USER-''
prefixes were eliminated; b) the ``USER-MISC'' package that had become
very large was split into several ``EXTRA-'' packages and some core
styles were moved there as well; c) packages with similar applications
would have the same prefix (e.g. ``DPD-'' for dissipative particle
dynamics variants, ``CG-'' for coarse-grain models, or ``ML-'' for
machine learning potentials).

\subsection{Removing and Replacing Packages, Tools, and Functionality}
\label{sec:refactor:remove}

The most drastic refactoring changes are the removal of entire packages
and bundled tools (Table~\ref{tab:removed}) and the reimplementation of
selected functionality.  Most often a package was removed only after one
with equivalent or improved functionality had been added and validated
against it.  The \texttt{REAX} package, a Fortran implementation of the
ReaxFF model\cite{vanDuin2001,vanDuin2003}, was replaced by
\texttt{USER-REAXC}, an adaptation of the C-based PuReMD
program\cite{puremd_paper}; the two coexisted until careful testing allowed
\texttt{REAX} to be removed and \texttt{USER-REAXC} to be renamed
\texttt{REAXFF}.  Likewise the Fortran \texttt{MEAM} package\cite{MEAM} was
re-implemented in C++ as \texttt{USER-MEAMC}---largely line by line, but
turning the Fortran global variables into class members so that multiple
instances can coexist---and renamed \texttt{MEAM} once validated as fully
compatible.

For other packages the functionality was preserved but realized
differently.  The \texttt{MESSAGE} and \texttt{LATTE} packages, which
coupled LAMMPS to external simulation software\cite{latte}, were both
superseded by the more general \texttt{MDI} package built on the MDI
library\cite{mdi_home}.  The \texttt{USER-CUDA} package was not replaced in
kind but served as the base of the \texttt{KOKKOS}
package\cite{lammps_kokkos,kokkos22}, which targets a variety of accelerator
hardware rather than only CUDA.  Finally, some packages were removed outright
because their code had been unmaintained for a long time and either carried
known bugs or broke with modern C++ compilers: this applied to \texttt{MPIIO}
and \texttt{MSCG}, and most recently to \texttt{ATC}, \texttt{AWPMD}, and
\texttt{POEMS}.  The removal of the latter three is clearly visible as a drop
in the Coverity Scan defect count (Fig.~\ref{fig:scan_defects}), confirming
that those packages were a significant source of issues reported by static
code analysis (Sect.~\ref{sec:test:static}).

The bundled tools shipped with LAMMPS are generally less well maintained
than LAMMPS itself and were likewise removed or replaced over time.  The
\texttt{restart2data} converter for turning nonportable binary restart files
into portable ``data'' files became tedious to keep in sync with the core
code and was replaced by a command-line flag that internally runs
\texttt{read\_restart} followed by \texttt{write\_data}.  The minimal X11
trajectory viewer \texttt{xmovie} was dropped rather than ported to 64-bit
platforms, since tools such as
VMD\cite{vmd_96,vmd_home,topotools_paper,topotools_code} and
OVITO\cite{ovito_20,ovito_home} support LAMMPS files and are more than
adequate replacements.  The obsolete Python\,2 script
\texttt{amber2lammps.py} was replaced by a Howto document for the external
AMBER2LAMMPS\cite{amber2lammps} tool.  The LAMMPS-SHELL program was removed
once it had been superseded by LAMMPS-GUI\cite{lammps_gui_home}; the
LAMMPS-GUI sources, the moltemplate tool\cite{moltemplate}, and the i-PI
package\cite{ceriotti14,Litman24,i_pi}---which uses LAMMPS as a force engine
for path-integral simulations through the \texttt{fix ipi} command---were in
turn unbundled because they had grown popular enough to be maintained as
separate projects.

In a few cases entire pieces of functionality were reimplemented rather than
merely cleaned up, either to remove an external dependency or to replace a
narrow feature with a more general one.  A recurring goal has been to
eliminate LAMMPS' dependence on Fortran---and thus on a Fortran compiler,
which is not always available on Windows.  Packages that relied on an
external BLAS/LAPACK library now fall back to the subset of the reference
(netlib) routines\cite{lapack} that LAMMPS actually uses, converted to C++
almost entirely with the \texttt{f2c} tool\cite{f2c} (only a few routines
were fixed by hand) and now maintained as a separate
project\cite{lammps_linalg}; the Fortran-to-C++ conversion of \texttt{REAX}
and \texttt{MEAM} above served the same goal by a different route.  A second
kind of replacement consolidates several narrow commands into one general
framework: the \texttt{fix ave/spatial} and \texttt{fix ave/spatial/sphere}
commands were superseded by the combination of \texttt{compute chunk/atom},
which assigns atoms to arbitrary ``chunks'' by region, molecule, type, or
computed property, and \texttt{fix ave/chunk}, which averages over them---a
more flexible and extensible design, since new ways of defining chunks and of
reducing per-chunk data can be added independently.

\subsection{Adoption of Modern C++ Features}
\label{sec:refactor:modernize}

As the C++ standard evolves over time, the features of the C++ standard
library become more reliable and powerful.  The LAMMPS developers have
thus decided to gradually abandon the strict ``C with classes'' programming
styles.  This applies specifically to the \texttt{std::string} class
and STL container classes, particularly \texttt{std::vector}.  Eventually,
the minimum required C++ standard was raised to C++11 and, more recently,
to C++17 (see the timeline in Sect.~\ref{sec:history:cplusplus}).  C++11 was
a major upgrade over the previous C++98 standard and added much useful and
convenient functionality, and C++17 provides further opportunities to
modernize and improve LAMMPS.

Using standardized C++ features eliminates the need to have custom
implementations or platform abstractions in LAMMPS (Sect.~\ref{sec:refactor:namespaces})
for compilers and platforms that are fully standard compliant.
For example, the \texttt{std::filesystem} namespace in C++17 replaces
some of the platform abstractions already in LAMMPS.  With the
``C with classes'' programming style, LAMMPS depends on the equivalent
functionality of the standard C library.  However, that is not as portable
and consistent since only a part of it is standardized in the C-language
standards, other parts conform to standards like POSIX which are not
universally available or use different conventions on different platforms.

There are a few exceptions where the original code is retained for
simplicity or efficiency reasons.  For example, LAMMPS continues to use
the ``stdio'' library for file I/O instead of the more complex and less
efficient ``iostreams'' of C++.  Furthermore, for large memory
allocations (for example per-atom vectors or arrays, or neighbor lists),
LAMMPS employs wrapper classes calling \texttt{malloc()} and
\texttt{free()} to reduce overhead from memory management and memory
fragmentation, lower the memory use, improve CPU cache efficiency, and
simplify MPI communication.

The modernization changes are too many to discuss them all in detail
here.  Instead, we will focus on a few exemplary changes.  The overarching
goal is to retain or---where possible---improve on a key property of
the LAMMPS source code: it is easy to extend and modify in order to add
new models, system manipulations, and on-the-fly analysis computations.
The more complex and sophisticated source code is restricted to parts
that usually do not require changes for adding new functionality.

\subsubsection{String Handling and Formatting}

C-style string handling is rather complex, error-prone, and difficult to
implement in a safe manner because of having to adjust dynamically
allocated memory all the time and handling pointers.  Thus, the
\texttt{std::string} class is used in many places, especially for local,
temporary strings, but also when passing strings to functions.  This is
aided by automatic conversions from \texttt{const char *} to
\texttt{const std::string \&}.  This and the \texttt{c\_str()} member
function of C++ strings to return a C-style string pointer have allowed
the code to be transformed to modern string handling incrementally.
However, some key parts will require a larger refactoring effort to be
transformed and therefore have been skipped so far.  The preference for
C++ strings is also reflected in the collection of functions in the
added \texttt{utils} namespace, as discussed below.

\label{sec:refactor:fmt}%
The preference for C++ strings and the adoption of C++11 is further
supported by integrating the \{fmt\} formatting library\cite{libfmt}.
It improves on the C-style \texttt{\*printf()} functions in being
type safe with a generic placeholder (\{\}) for any type of (supported)
argument.  The \{fmt\} library is predominantly used three ways: a) as
the \texttt{fmt::format} function directly to produce formatted strings
similar to \texttt{sprintf()}, b) through utility functions like
\texttt{utils::print()} which can be used like \texttt{fprintf()}
only that it uses a \{fmt\}-style format string, and c) as overload
to several functions that would previously only take a single string
as argument, but now may use a format string and a variable number of
arguments.  For example, this has massively simplified the output of
customized error messages.  Previously, this required allocating a
suitable-sized string buffer, using \texttt{snprintf()} to create the
custom string, output of the buffer, and freeing the temporary storage.
Now, this operation has become a single statement that differs only from
the static string version by having a variable number of additional arguments
(Listing~\ref{lst:fmt}).

\begin{lstlisting}[float=tb,caption={Building a custom error message before and after adopting the \{fmt\} library: the manual buffer handling (top) collapses into a single, type-safe call (bottom).},label={lst:fmt}]
// before: manual C-style buffer handling
char *buf = new char[128];
snprintf(buf, 128, "bad value %g", cutoff);
error->all(FLERR, buf);
delete[] buf;

// after: type-safe {fmt}-style formatting
error->all(FLERR, "bad value {}", cutoff);
\end{lstlisting}

Many scientists are familiar with the Python programming language,
so the similarity of \{fmt\} formatting to Python string formatting is
welcome.  The \{fmt\} formatting functions have since been adopted into the
C++20 standard as \texttt{std::format}.  We have taken the next step and
added wrapper headers so that, when LAMMPS is compiled with C++20 support
and \texttt{std::format} is available, the \texttt{fmt::format} calls are
mapped onto \texttt{std::format} and the bundled copy of the \{fmt\} library
is no longer required.  Until C++20 can be relied upon on all platforms that
LAMMPS supports, the bundled \{fmt\} implementation remains the fallback.

\subsubsection{Utility Functions and Classes}
\label{sec:refactor:namespaces}

As with any large software package, there are repetitive tasks that can
be abstracted by implementing custom functions that perform those tasks.
Not only does this reduce the amount of repeated code, it improves
consistency and avoids having to fix the same bug in multiple locations.
Some of these functions were originally implemented as members of the
foundational LAMMPS classes. During the modernization these functions
were moved to a \texttt{utils} namespace, and then many more utility
functions were added.  These provide tasks like transforming strings (to
upper or lower case, from UTF-8 to ASCII), removing comments, trimming
or compressing whitespace, splitting or joining strings following
LAMMPS-specific rules, or conversion of strings to numbers and many
more. These functions have extensive unit tests
(Sect.~\ref{sec:test:unittest}).

Similarly, platform-specific functionality was collected and abstracted
into the \texttt{platform} namespace.  As a consequence, the LAMMPS code
has very few places where C pre-processing is required for conditional
compilation on different host platforms.  One such abstracted feature
is the traversal of paths and folders and the manipulation of filenames.
Another abstracted platform-specific feature is the dynamic loading of
object files as part of the ``plugin'' command
(Sect.~\ref{sec:build:plugins}).

\label{sec:refactor:tokenizer}%
Some of these repetitive tasks are better handled by specialized classes.
Most prominent are the
\texttt{Tokenizer} and \texttt{FileReader} classes and their variants.
Many of LAMMPS' pair styles require reading and parsing of files with
the potential parameters.  Historically, this was done using C-style
string buffers and the \texttt{strtok()} function. This function is not
reentrant and overall its use led to rather complex and hard to read
(and debug!) code.  Using the new tokenizer and file reader classes
reduced the number of lines of code required to read and parse parameter
files by a factor of 3--5, sometimes even more, and made the resulting
code more readable (Listing~\ref{lst:tokenizer}).

\begin{lstlisting}[float=tb,caption={Parsing a line of potential
 parameters before and after the \texttt{Tokenizer} classes.
 The non-reentrant \texttt{strtok()}/\texttt{atof()} idiom (top)
 needs a manual check for every field and silently yields zero
 on malformed input, whereas a \texttt{ValueTokenizer} (bottom)
 extracts and type-checks each value and throws an exception
 (Sect.~\ref{sec:refactor:exceptions}) on error.},label={lst:tokenizer}]
// before: not reentrant; each value needs a manual check
char *word = strtok(line, " \t\n");
if (!word) error->all(FLERR, "missing element name");
std::string iname = word;
word = strtok(nullptr, " \t\n");
if (!word) error->all(FLERR, "missing epsilon");
// silently 0.0 if malformed
double epsilon = atof(word);

// after: reentrant and each numerical value is validated
ValueTokenizer values(line);
std::string iname = values.next_string();
// throws on bad input
double epsilon    = values.next_double();
\end{lstlisting}

\subsubsection{Exceptions and Resource Management}
\label{sec:refactor:exceptions}

Using C++ exceptions improves the flexibility of error handling and
simplifies code in many places.  This has been introduced locally in
places, e.g., when using the tokenizer classes
(Sect.~\ref{sec:refactor:tokenizer}). But this has also been applied to
the general error handling in the \texttt{LAMMPS\_NS::Error}
class. Rather than aborting the LAMMPS simulation directly, an exception
is thrown and caught by an exception handler in the \texttt{main()}
function. This exception handler also processes exceptions thrown by C++
standard library functions and classes or the \{fmt\} library and thus
makes it easier to determine the cause of an error.

An important benefit exists for the C-language library interface
(Sect.~\ref{sec:library}) where all functions that
may potentially throw exceptions are wrapped into an exception
handler, and functions were added to detect if an error has
occurred and to retrieve the error message.  This prevents
programs using the library interface from crashing immediately;
instead, they can check the status and then choose how to proceed,
that is, either ignore the error or recreate the LAMMPS instance
and repeat the simulation with different settings.  The same mechanism is
exploited by the LAMMPS-GUI application (Sect.~\ref{sec:gui}), which runs
LAMMPS in a separate thread: because an error throws an exception rather than
aborting the process, the GUI can report it and let the user correct the
input instead of crashing.

\label{sec:refactor:raii}%
\begin{lstlisting}[float=tb,caption={Closing a possibly compressed file before and after \texttt{SafeFilePtr}.  The manual version must track whether the stream was opened with \texttt{popen()} and close it on every code path; forgetting \texttt{pclose()} closes the stream but leaks the pipe's file descriptor.  \texttt{SafeFilePtr} records this once and closes the file with the correct call automatically when it goes out of scope.},label={lst:raii}]
// before: track pipe vs. file, close on every path
FILE *fp;
bool pipe = false;
if (has_compress_extension(file)) {
  fp = compressed_read(file);  // uses popen()
  pipe = true;
} else {
  fp = fopen(file, "r");
}
// ... read the file ...
if (pipe) pclose(fp); else fclose(fp);

// after: closed with the correct call at scope exit
SafeFilePtr fp;
if (has_compress_extension(file)) {
  fp.set_pclose();
  fp = compressed_read(file);
} else {
  fp = fopen(file, "r");
}
\end{lstlisting}

A further modernization is the gradual adoption of the RAII idiom
(``resource acquisition is initialization''), in which ownership of a
resource is tied to the lifetime of an object so that the resource is
released automatically when that object goes out of scope.  This pairs
naturally with the use of exceptions (Sect.~\ref{sec:refactor:exceptions}),
since it guarantees that resources are released even when the stack is
unwound by an error.

A representative example is the \texttt{SafeFilePtr} class, a drop-in
replacement for a \texttt{FILE *} variable that closes the file
automatically when the variable goes out of scope or is reassigned.  This
both simplifies the code and removes a class of bugs: many file reading and
parsing routines in LAMMPS have complex loop and branch structures that would
otherwise require a matching \texttt{fclose()} on every code path, which is
easy to get wrong and also tends to confuse static analysis tools into
reporting spurious file-pointer leaks.  In addition, a \texttt{set\_pclose()}
call records that a file was opened with \texttt{popen()}---as is done for
reading and writing compressed files---so that the matching \texttt{pclose()}
is used when the file is closed.  Previously, such cases called only
\texttt{fclose()}, which closed the stream but leaked the file descriptor of
the pipe (Listing~\ref{lst:raii}).  Such a leak is harmless in a short
standalone run that the operating system cleans up on exit, and so it had
long been tolerated as a minor issue; but it becomes a real problem in
long-running workflows driven through the Python module
(Sect.~\ref{sec:library:python}), where many files are opened and many
LAMMPS instances are created and destroyed within a single process, so that
leaked descriptors accumulate until the per-process limit is reached.

A second example is the use of \texttt{std::vector} for temporary local
allocations in place of \texttt{new[]}/\texttt{delete[]} or the LAMMPS
\texttt{memory->create()}/\texttt{memory->destroy()} helpers.  A vector frees
its storage automatically and is almost a drop-in replacement; only access to
the underlying raw pointer requires an explicit \texttt{.data()} call.  This
is especially useful for replacing variable-length arrays, which are allowed
in Fortran and in some C dialects but are \emph{not} part of the C++ standard.
Because several open-source C++ compilers accept them as an extension, such
non-portable code can otherwise go unnoticed---particularly when written by
contributors with a background in Fortran.

A third example is the use of \texttt{std::unique\_ptr}, typically created
with \texttt{std::make\_unique}, for objects that are allocated on the heap
but owned by a single class or scope.  The object is then deleted
automatically when its owner goes out of scope, is reassigned, or is
destroyed; this removes the explicit \texttt{delete} calls and the leaks that
occur when one is forgotten on an early return or while an exception is
propagating.  For example, \texttt{fix bond/react} holds its optional JSON
reaction-metadata object in a \texttt{std::unique\_ptr} member that is
created on demand with \texttt{std::make\_unique}, and the \texttt{KSPACE}
FFT wrapper holds its FFT plan in a \texttt{std::unique\_ptr} member; in both
cases the class no longer needs an explicit \texttt{delete} in its destructor.

\subsubsection{Accessors and Error Messages}
\label{sec:refactor:accessor}

As a consequence of LAMMPS being designed initially in Fortran,
internal data structures or settings of classes were accessed
directly when it was translated to C++.  Fortran at the time did
not have a feature equivalent to classes, and best practices of
object-oriented program design were not as well understood as they
are now.  This direct access can lead to problems
when the internal state is changed and the settings depend on each
other.  There is a risk that the settings are not updated consistently
because the programmer does not know about the dependency or
a feature was implemented when the dependency did not exist.
Similarly, having to know the details of the internal data structures
can make the code more complex compared to the cases where STL containers
are used to return data.

Thus, a part of the ongoing refactoring effort replaces direct access
to the foundational classes of a LAMMPS simulation with accessor
functions that return either a pointer or a list of pointers to
requested objects. Other added functions
are used to set the internal state of a class according to some
flag and then enforce the consistent setting of all dependent flags.
These accessor functions often also reduce repetitive code when
a loop over elements of a list is implemented in the accessor
instead of the code that calls it.

Overall, improving the separation between the internal state
of core LAMMPS classes and the functionality programmed by LAMMPS
users and code contributors helps to improve the maintainability
of LAMMPS. Either part can be developed independently for as
long as the interface is preserved.

\begin{figure*}[tb]
\begin{lstlisting}[language={}]
ERROR: Unrecognized fix style 'ilves' (src/modify.cpp:935)
Last input line: fix 0 all ilves 0.0001 100 1 b 1 2 a 1
                           ^^^^^

ERROR: Unknown identifier in data file: Massess
For more information see https://docs.lammps.org/err0001 (src/read_data.cpp:1482)
Last input line: read_data       data.peptide
\end{lstlisting}
\caption{Two example LAMMPS error messages.  The first underlines the offending command argument with a row of caret characters; the second appends a coded URL (produced by \texttt{utils::errorurl()}) that points to a manual page with a detailed explanation.  Both echo the input line that triggered the error together with the source file and line where it was raised.}
\label{lst:error}
\end{figure*}

\label{sec:refactor:errors}%
Another refactoring effort concerns the content of error messages
that LAMMPS would issue.  Historically, many error messages were
of the kind ``there was an error'' with no details about the
cause and how to remedy it. Removing the error would require a
careful study of the manual and often also the source code to
determine the exact cause. The latter part was made easier by
adding a pre-processor macro so that the error message would
contain the name of the source file and the line number where
the error message was dispatched.  However, this became a growing
problem as the audience of LAMMPS shifted from scientists familiar
with programming and able to read source code to ones that use
pre-compiled executables and have little programming knowledge,
and in some cases no direct access to the specific source code
that was used to compile the LAMMPS executable they are using.

We use the following three steps to improve the error message:
a) we expand error messages to state which keyword was failing,
and thanks to including the \{fmt\} library (Sect.~\ref{sec:refactor:fmt})
error messages can include more details about the faulty data;
b) for errors requiring a more detailed explanation, we add
a mechanism to output a coded URL that refers to a specific
section in the LAMMPS online manual with a discussion of that
particular error;
c) where possible, we repeat the offending line of input as part of the
error message and use ASCII graphics to mark the specific word that caused
the error.

This last mechanism is selected through a second form of the error functions
that takes an integer argument between the \texttt{FLERR} macro and the
format string, as in \texttt{error->all(FLERR, idx, \dots)}.  Here
\texttt{idx} is the index (starting at zero) of the command argument that
failed, so that the corresponding word can be marked; the special constant
\texttt{Error::COMMAND} points at the command name instead.  When a variable
was substituted in the offending line, the message shows the line both before
and after substitution, since the cause is often apparent in only one of the
two.  Further constants tune the behavior: \texttt{Error::NOPOINTER} prints
the line without a marker, while \texttt{Error::NOLASTLINE} omits the echoed
line altogether---preferred for errors raised \emph{during} a run or
minimization, where the last processed line is the unrelated \texttt{run} or
\texttt{minimize} command.  Fig.~\ref{lst:error} shows two representative
examples; these conventions are documented in the ``Programmer's Guide''
(Sect.~\ref{sec:docs:developer}).

\section{Library Interfaces and Language Bindings}
\label{sec:library}

An unusual feature of LAMMPS is that it is designed so that
an instance of the \texttt{LAMMPS\_NS::LAMMPS} class holds
the complete state of a simulation.  The \texttt{main()}
function of the LAMMPS executable is quite minimal and mostly
creates such a class instance while passing the command-line
arguments and then sequentially processes each line of input
until it reaches the end.  This enables, for example, the use
of LAMMPS as a component in multiscale applications\cite{library_interface}
or in task farming workflows.  Especially the LAMMPS Python
interface (Sect.~\ref{sec:library:python}) has become
very popular for these types of applications.
This is not limited to a single LAMMPS instance, but custom
applications can be written that create multiple class instances,
either sequentially or concurrently.\footnote{LAMMPS is not fully
re-entrant and thus running multiple LAMMPS instances at the
same time requires using separate processes on separate MPI
communicators that were split off the MPI world communicator.
Alternating between multiple LAMMPS instances in the same process
is possible; it is still recommended to create a separate MPI
communicator for each instance.}

\subsection{C-language Interface}
\label{sec:library:c}

To make the encapsulation of a simulation more accessible,
a C-language library interface was added, where only a single
header \texttt{library.h} would be needed to have access to
functions that create, query, and manipulate a LAMMPS class
instance.  Since the interface uses C and not C++, it is
rather straightforward to call it from other programming languages,
either directly or by providing an interface file for
the SWIG tool\cite{swig}.

Historically, the LAMMPS library interface was created in a
rather \textsl{ad hoc} fashion, that is, new functions were added
when someone needed them, and the specific use case would determine
the type and number of arguments of those functions.  This led
to a rather inconsistent design and some degree of redundancy.
Since removing inconsistencies and redundancies would result in
some incompatible changes, the decision was made to have a major
overhaul of the library interface that would make future
breaking changes unlikely.

Additional motivation for a refactoring was the inconsistent
documentation (Sect.~\ref{sec:docs:doxygen}) and the need
for a cleaner and more consistent Python interface
(Sect.~\ref{sec:library:python}). The library interface
is also used in most of the unit test tools
(Sect.~\ref{sec:test:unittest}) and, in
combination with the Python interface, a significant
expansion of the introspection abilities was needed.

\subsection{Python Interface}
\label{sec:library:python}

The LAMMPS Python interface is a Python package that provides
access to the C-language library interface (Sect.~\ref{sec:library:c})
from Python using the ``ctypes'' package to interface with
dynamically loaded libraries; this
requires building LAMMPS as a shared library instead of the default
static library.  Initially, it followed the same \textsl{ad hoc}
development model as the C-language interface and, as a consequence,
was neither consistent nor complete. Thus, it was refactored and made
more consistent alongside the refactoring of the C-language
interface. However, it also required extensions to the C-language
interface, especially introspection functions, which would then
be used to make the behavior more Python-like. That includes the
use of exceptions (Sect.~\ref{sec:refactor:exceptions}) which are
monitored by the Python interface and then re-thrown as Python
exceptions, if needed.

Unlike the C-language interface, which due to the restrictions
of the C programming language, consists of a collection of global
functions that use a common \texttt{lammps\_} prefix,
and where the first argument is an opaque pointer to the instance
of the \texttt{LAMMPS\_NS::LAMMPS} class, the Python interface uses
a \texttt{lammps} Python class, with the functions of the library
interface becoming methods of the class.

In addition, new functionality was added to the LAMMPS Python interface,
such as accessing arrays via NumPy\cite{numpy,numpy_web}, a \texttt{cmd} property that
makes code look more ``pythonic'' (example: \texttt{lmp.command("units lj")}
becomes \texttt{lmp.cmd.units("lj")}; see Listing~\ref{lst:python}), an
\texttt{ipython} property that integrates the Python \texttt{lammps} class
with Jupyter notebooks (e.g. to embed snapshot images created by the
\texttt{dump image} command).

\begin{lstlisting}[language=Python,float=tb,caption={Driving LAMMPS from Python.  Arbitrary commands can be issued as strings through \texttt{command()}, exactly as in a LAMMPS input script, or equivalently through the \texttt{cmd} property, which turns each LAMMPS command into a method call; per-atom data is exposed directly as NumPy arrays without copying.},label={lst:python}]
from lammps import lammps
lmp = lammps()

# command strings, exactly as in a LAMMPS input script
lmp.command("units lj")
lmp.command("region box block 0 10 0 10 0 10")

# the same commands through the pythonic cmd property
lmp.cmd.units("lj")
lmp.cmd.region("box block", 0, 10, 0, 10, 0, 10)

# per-atom positions as a NumPy array (no copy)
x = lmp.numpy.extract_atom("x")
\end{lstlisting}

Finally, the initial single-file Python ``wrapper'' module was refactored
into a regular Python package with a corresponding directory structure.
The installation support based on the deprecated ``distutils'' was updated
to use ``setuptools'' instead. The build process was structured such
that an ``install-python'' target was added to the build system which
executes the build of the LAMMPS Python package with ``pip'', ``build'',
``wheel'', and ``setuptools'' into a so-called wheel archive which
contains not only the Python code, but also the LAMMPS shared library.
This makes the ``wheel'' self-contained and directly installable through
the \texttt{pip} command.  In fact, the ``install-python'' build target
will attempt to install the created wheel of the \texttt{lammps} Python
package.

\subsection{Fortran Interface}
\label{sec:library:fortran}

Several attempts at providing a Fortran interface to LAMMPS have been
made over the years, and their source code has been included into the
LAMMPS source code distribution. But they were incomplete and most
relied on some additional library of Fortran functions to support the
interface.

In 2020, the development of a new Fortran interface was started,
reusing as much of the existing interface code as possible, but with
the following design decisions: a) the entire interface is contained
in a single source file providing a \texttt{LIBLAMMPS} Fortran module;
b) the module provides a \texttt{LAMMPS} derived type that can be
used in ways very similar to the \texttt{lammps} class of the
LAMMPS Python interface; c) a compiler compliant with Fortran 03 is
required since the ISO\_C\_BINDINGS module is used for calling the
functions in the C-language library interface;
d) the module itself does not contain any MPI library calls and
the optional communicator argument of the \texttt{lammps} class uses
Fortran 77 style MPI communicators that are represented by integers.
Internally, those can be inter-converted using the \texttt{MPI\_Comm\_f2c()}
and \texttt{MPI\_Comm\_c2f()} functions.  To make d) work, a
\texttt{lammps\_open\_fortran()} ``constructor'' function was added
to the C-language library interface to support passing an MPI 
communicator through its integer representation.

Thanks to the help of contributors of previous Fortran interfaces,
the new LAMMPS Fortran now interfaces with the complete C-language
API and is integrated into the automated testing procedure (Sect.~\ref{sec:test:unittest}).
All previous interfaces have been declared obsolete and removed
from the distribution. Thanks to the single source file approach,
adding this new interface to a Fortran code requires adding this
one file to the build process so it gets compiled before any
files that make use of the module and link to the LAMMPS library,
by preference with the shared library so that all dependencies
to external libraries of LAMMPS itself are automatically included.

\section{Build System and Deployment}
\label{sec:build}

The way LAMMPS is built, packaged, and distributed has changed as much as
its source code.  Two developments stand out: the gradual replacement of the
traditional GNU Make build with a CMake-based system that also enables
ready-to-use binary packages for all major platforms, and a plugin mechanism
that lets external packages be added to a LAMMPS executable at runtime
without rebuilding it.

\subsection{CMake-based Build System}
\label{sec:build:cmake}

The traditional build system of LAMMPS was designed around the
GNU Make program in combination with some Bourne shell and
Python scripts to handle package dependencies and specific
build requirements. The build process consists of three
steps: 1) building bundled libraries in the \texttt{lib} sub-folders
or interfacing to externally built libraries, and 2) enabling
or disabling packages (by copying files from package folders
to the main source folder or by deleting them from the main
source folder) and 3) compiling the sources in the main source
folder into a LAMMPS library and executable based on so-called
``machine makefiles'' that were customized for specific compilers,
operating systems, features and libraries, and computer hardware.
Over the years, this process has been optimized and sources of
errors reduced, but it intrinsically requires a good understanding
of using compilers and libraries, and adjustments have to be made
manually. During this time, asking for help compiling
LAMMPS was a recurring topic of discussion on the LAMMPS mailing list.

After intense discussions, a concerted effort was started
in August 2017 to implement an alternative build system based on
the CMake tool\cite{cmake}.  Original design goals were that the behavior of
the build process, specifically how to include optional packages,
should be similar and that both build systems should co-exist.
CMake offers several advantages: it can largely automate the detection
and configuration of features based on available tools and libraries
and the operating systems; it supports out-of-source builds (in fact,
LAMMPS' CMake scripting enforces that) and thus separates binaries and
sources and allows one to build different configurations from the exact
same source code in the same tree; it provides a platform neutral and
platform aware script language that simplifies porting of LAMMPS to
different operating systems and compiler environments; it provides a
build and installation process similar to many other software packages,
and thus simplifies creation of pre-built binary packages for Linux
distributions and automated build environments like Homebrew\cite{homebrew}
or Spack\cite{spack}.
Over the years, the CMake build process has been continuously improved
and modernized and includes features that are not available to the
legacy build process, such as unit testing (Sect.~\ref{sec:test:unittest}).

Specifically, the addition of CMake support has enabled compiling and developing
LAMMPS on Windows with Microsoft Visual Studio\cite{MSVS_community},
where previously the installation of a Unix-like environment for Windows (like
MinGW or Cygwin) was required; a step that only a few Windows users were willing
to do in the past.  Furthermore, it enabled creating easy-to-install packages
with pre-compiled binaries of LAMMPS and LAMMPS-GUI for Windows (as an installer
package), macOS (as \texttt{dmg} disk image file), and Linux with x86\_64 CPUs
(as tar archive with wrapper scripts and, more recently, as Flatpak\cite{flatpak}
 bundles).

Starting with the 10 September 2025 release the legacy build process no longer
supports any packages that require the building of or interfacing with
libraries in the \texttt{lib} folder, which is the most error-prone step of
the legacy build system.  The expectation is that eventually only the CMake build
system remains.  Having only the CMake build system will facilitate further improvements
in the build process that are currently prevented by having to support the
legacy build process as well.

\subsection{Plugins}
\label{sec:build:plugins}

Due to the popularity of LAMMPS and the ease with which it can be
modified and extended, there are many projects with add-on software
for LAMMPS that are maintained separately.  For these packages, it
can be difficult to integrate them cleanly into the LAMMPS
build procedure. Many such projects include patch files with modifications
to one or both of the LAMMPS build systems.  Due to the ongoing
development of LAMMPS, these patches may break when applied to a
different LAMMPS version than what was used by the developers of
the add-on software.

We have thus developed a plugin mechanism that would allow developers
and users of external packages to build the add-on functionality as
a separate shared object that would be dynamically loaded into the
LAMMPS executable at runtime with the ``plugin'' command. This
mechanism also enables distributing binary packages as
plugins for LAMMPS where the add-on package or a library upon
which it depends would be incompatible with the LAMMPS license (GPLv2).

\section{Improved Documentation}
\label{sec:docs}

\begin{figure*}[tb]
\centering
\begin{minipage}[b]{0.48\textwidth}\centering
  \includegraphics[width=\linewidth]{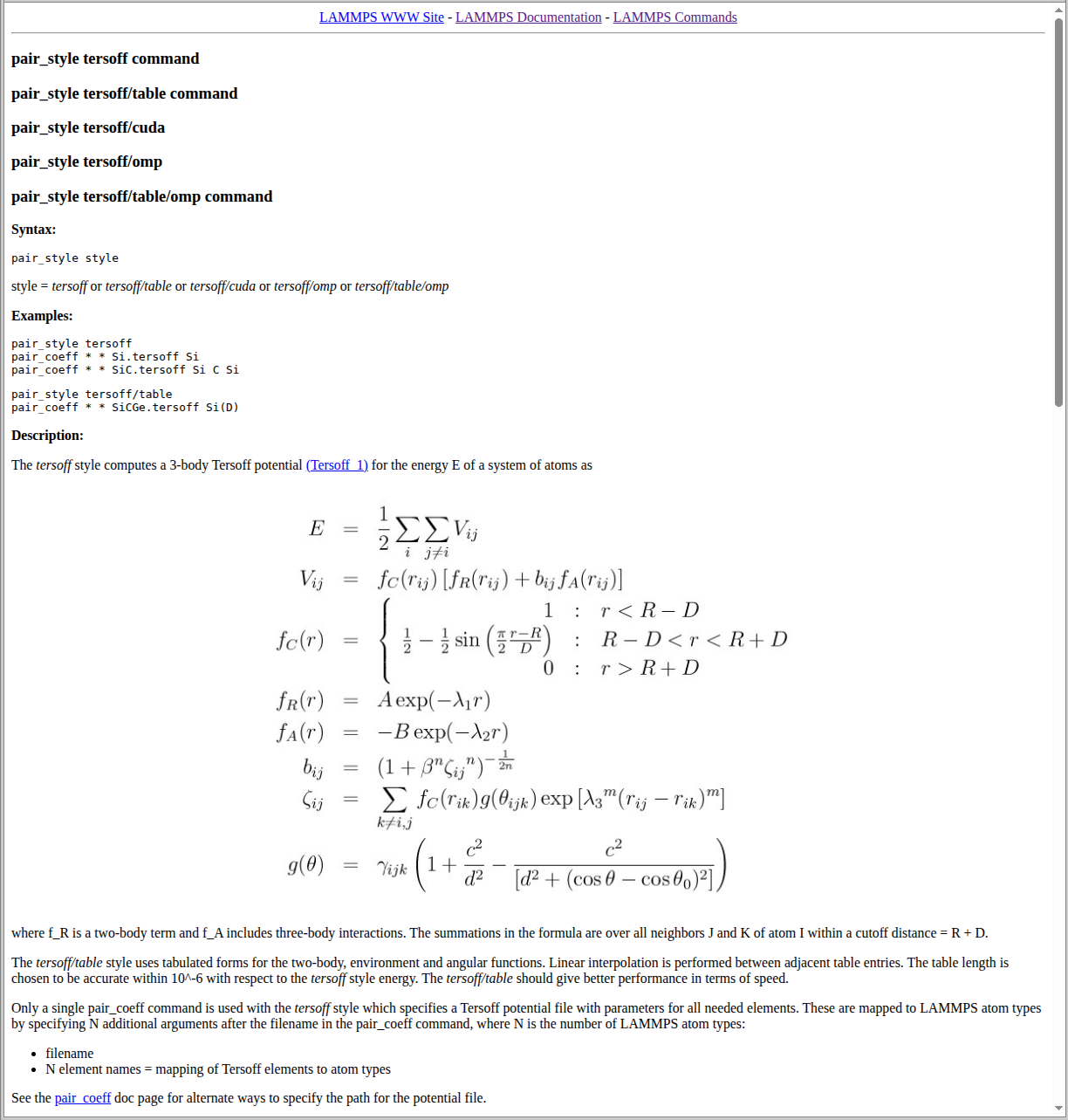}
\end{minipage}\hfill
\begin{minipage}[b]{0.48\textwidth}\centering
  \includegraphics[width=\linewidth]{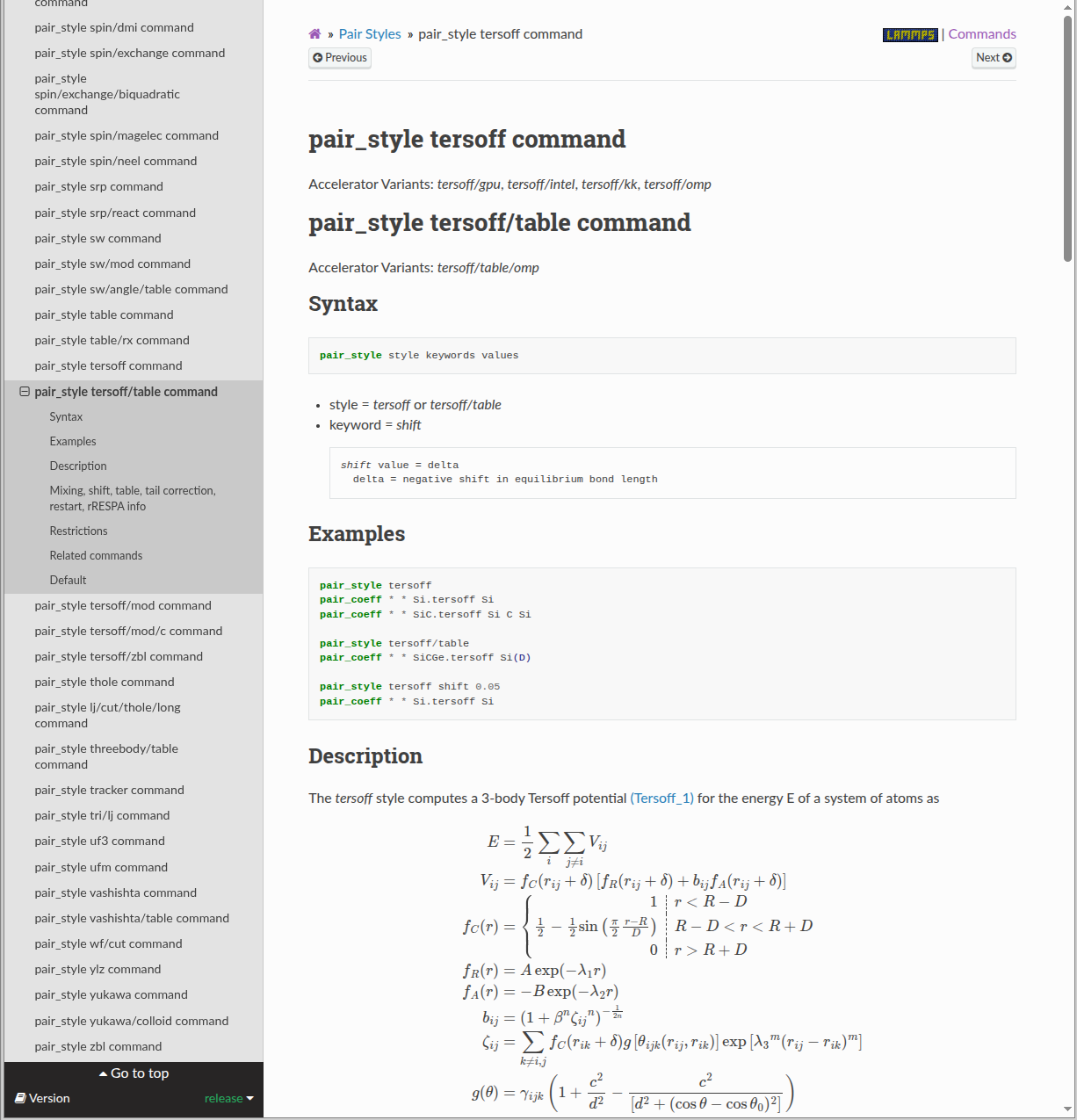}
\end{minipage}
\caption{The manual page for the Tersoff potential as rendered by the
  original home-grown HTML generator (left) and by the current Sphinx
  and reStructuredText based documentation (right), which adds
  navigation, equations typeset with MathJax instead of images, and
  syntax-highlighted input examples.}
\label{fig:manual}
\end{figure*}

In step with the modernization of the LAMMPS source code, we are modernizing
the documentation.  Historically, LAMMPS used a simple, home-grown markup
language that would be translated into HTML format on a file-by-file basis.
This was simple and very fast, but had a major drawback: mathematical expressions
had to be rendered into images first.  The resulting HTML looked rather plain
and reflected the typical style of web pages in the 1990s and early 2000s.
In addition, navigation was minimal. The PDF version of the manual,
including its table of contents, was created directly from the generated HTML
pages with a tool called HTMLDOC\cite{htmldoc}.  Figure~\ref{fig:manual}
contrasts the old and the current rendering of the same manual page.

Due to the large number of features and packages that were added over
the years, the LAMMPS documentation has become large: currently (2026
stable version) it consists of over 1050 source files in
reStructuredText format (not counting documentation that is extracted
from C++ or Python source code) and has over 3200 pages when translated
into PDF format.

\subsection{Use ReStructuredText and Sphinx}
\label{sec:docs:sphinx}

Initially, the simple markup was retained and then converted on the fly with
a custom tool to reStructuredText\cite{rst}. Those \texttt{rst} files would
be translated into HTML or PDF with the Sphinx\cite{sphinx} tool.
This required some ``massaging'' of some of the documentation sources to
obtain a readable translation, and some minor flaws were unavoidable.
However, the output was much more pleasant to look at.

In a second step, the translated \texttt{rst} files became the canonical
source files for the documentation, and any new documentation files
would be added only in reStructuredText format.  With the available
translation tools, it would still be possible for veteran LAMMPS
developers to write a first draft in the old markup, translate that into
\texttt{rst}, and correct and polish the \texttt{rst} file for
submission.  Over time, the number of cross-references to other commands
or specific sections of the manual has been growing, thus providing
readers of the manual with more context.

Once the \texttt{rst} files became the canonical source, further
improvements could be added, including some provided by Sphinx
extensions and others from custom additions. Notable is, for example,
the use of MathJax\cite{mathjax} for typesetting mathematical
expressions in \LaTeX{} so that equations are no longer imported as
images, which makes adding and correcting them much easier.  Or, the use
of syntax highlighting for code examples, including LAMMPS input
examples.

Custom scripts were added to ensure that all included commands are
documented, included in the index, listed in the command summary tables
and lists, and that the markers for accelerated versions are correct. In
addition, references are verified to be available and unique.

\subsection{Integrate Python Docstrings and Doxygen Comments}
\label{sec:docs:doxygen}

Switching to Sphinx and reStructuredText enabled a deeper integration between
source code and documentation by including Python docstrings and Doxygen-style\cite{doxygen}
comments into relevant parts of the manual.  Specifically for the library interface
this was a major improvement, since previously there were multiple versions of
descriptions: comments in the source code, a README file, some documentation in
the manual about the C-language version, and the Python wrapper that sits on top
of that.  Sadly, those were usually not in agreement with each other and with what
was implemented in the source code.  Having the canonical documentation embedded
as comments or strings into the source code resulted in just one location (with
cross-references) with the detailed description focused on the C-language
interface and others mostly describing what is different in their specific
language bindings compared to the C-language version.

\subsection{Adding a Programmer's Guide}
\label{sec:docs:developer}

The largest addition to the LAMMPS manual in recent years has been the
``Programmer's Guide'' part.  As we were modernizing the source code, we felt
the need to better document the various aspects of the LAMMPS program design
and its conventions and available abstract programming patterns.  This is done
with the understanding that most contributors to a scientific software package
are \emph{not} formally trained programmers, but scientists with the need to
extend or modify simulation software to realize their research projects.
Some parts were previously scattered across the manual, but having a dedicated
``Programmer's Guide'' section separates the documentation for users (which is
predominantly a command reference) from the documentation for programmers that
want to program LAMMPS or write programs on top of LAMMPS. 

This is a work in progress (and probably always will be for as long as
we keep refactoring and extending LAMMPS), but there is already a
substantial amount of information about a variety of topics available:
\begin{itemize}[noitemsep,topsep=0pt,parsep=0pt,partopsep=0pt]
\item[-] overall code design, flow of control during a run
\item[-] the core parallelization strategies;
\item[-] some high-level programming patterns;
\item[-] design of the polymorphic base classes for styles;
\item[-] the different library interfaces including commented examples of how to use them;
\item[-] utility or portability functions, classes, and tools;
\item[-] examples with detailed discussions about implementing some kinds of new styles from scratch;
\item[-] discussion of changed or removed features;
\item[-] porting source code written for older LAMMPS versions to work correctly with the current one;
\item[-] documentation of requirements and preferences when contributing code into the LAMMPS distribution.
\end{itemize}

\subsection{Maintenance Processes and Policies}
\label{sec:docs:policies}

The transformation from an ``autocratic'' development style with a
single main developer managing ``everything'' to a ``democratic''
style with a group of core developers making decisions and processing
contributions concurrently requires a formal description of the policies
and procedures governing the development and maintenance of LAMMPS.
This is provided in five markdown files: two in the \texttt{doc} folder
and three in the \texttt{.github} folder.

The documentation conventions file specifies the steps for adding
content to the documentation. The development workflow file specifies
the steps and policies for processing issues and pull requests on GitHub.
The release workflow file describes the steps to prepare a LAMMPS release
(feature, stable, or update). Finally, there are files stating the code
of conduct that must be followed and enforced by all developers, contributors,
and maintainers of LAMMPS and a general contributor's guide as recommended
by the GitHub best practices.

The availability of those documents is considered a way to simplify the
integration of new members to the team of core LAMMPS developers, as
well as a means of reducing the negative impact on the LAMMPS project
from the loss of core developers.

\section{Using AI in LAMMPS Development}
\label{sec:ai-usage}

\begin{table*}[tb]
\centering
\small
\setlength{\extrarowheight}{0.5ex}
\begin{tabular}{@{}>{\raggedright\arraybackslash}p{0.24\textwidth}p{0.335\textwidth}p{0.375\textwidth}@{}}
\toprule
Section and use case & Role in LAMMPS & Current assessment \\
\midrule
\ref{sec:ai-usage:mliap}:~Machine-learning interatomic potentials & ML models of interatomic interactions, usually trained on quantum-mechanical data. & Most mature use; near first-principles accuracy at much lower cost when well trained. \\
\ref{sec:ai-usage:input}:~Creating simulation inputs & Customized LLM chatbots that help non-experts assemble LAMMPS input scripts. & Still unreliable; works only for simple, tutorial-like cases; no model of the physics. \\
\ref{sec:ai-usage:review}:~Code review & Autonomous summaries and critical assessment of submitted pull requests on GitHub. & Uses project-specific instructions; catches formal and stylistic issues and pre-empts static analysis. \\
\ref{sec:ai-usage:debug}:~Debugging & Locating and fixing bugs from a comprehensive description and a reproducer. & Effective for well-documented, reproducible bugs; even failed attempts can help to find the cause. \\
\ref{sec:ai-usage:refactor}:~Refactoring & Modernizing code (and documentation) without changing behavior. & One of the strongest areas; sometimes edits need to be refined by an experienced developer. \\
\ref{sec:ai-usage:develop}:~Writing new code & Implementing new features from a plan and reference material. & Most demanding; can produce near-complete drafts given enough context and a frontier model. \\
\ref{sec:ai-usage:docs}:~Writing documentation and tutorials & Drafting and adapting manual pages, examples, and tutorial prose from code and notes. & Natural fit for language models; useful for routine drafting and language polishing, but every page must be verified against the code. \\
\bottomrule
\end{tabular}
\caption{Current AI usage in and around LAMMPS development, and our
  assessment as of mid-2026.  Details are in the indicated sections.}
\label{tab:ai-usage}
\end{table*}

In a scientific software package such as LAMMPS, the term ``artificial
intelligence'' (AI) covers several quite different things that are worth
distinguishing.  Machine learning has become an established
\emph{modeling} method in the form of machine-learning interatomic
potentials (MLIPs); assistants based on large language models (LLMs)
are increasingly used to help \emph{set up} simulations; and AI coding
agents are now applied to the \emph{development} of LAMMPS itself---for
code review, debugging, refactoring, and writing new code.  These uses
differ fundamentally in maturity.  MLIPs are a \emph{scientific}
application with a substantial body of published, quantitatively
validated results; we therefore treat them only briefly and point to
recent reviews (Sect.~\ref{sec:ai-usage:mliap}).

All of the other uses are \emph{software-engineering} applications that
have become practical only very recently and for which there is, as yet,
little established literature.  One observation recurs throughout: an AI
agent's output improves sharply with the amount and quality of context
it is given---a detailed plan, reference publications, template code,
and recorded lessons learned---so that the productivity gain comes not
from replacing an experienced developer but from amplifying one, who
must both assemble that context and judge the result.

What follows is therefore not a survey but a report of the LAMMPS
developers' own, still-ongoing first-hand evaluation: we state these
points explicitly as our current experience and assessment as of 2026
(Table~\ref{tab:ai-usage}).  Because the underlying tools, the available
models, and their cost evolve rapidly, they should be read as a snapshot
rather than a lasting assessment.

\subsection{Machine-learning Interatomic Potentials}
\label{sec:ai-usage:mliap}

For many computational scientists, machine-learning interatomic potentials
(MLIPs) are probably their first point of contact with AI in a molecular
dynamics context.  The boundary is gradual: fitting the parameters of a traditional
empirical force field to precomputed quantum-mechanical and experimental
reference data shares much of its methodology with training a
machine-learning model, and the two approaches form a continuum rather than
a sharp divide.  The most common MLIPs are trained on density-functional
data and promise results at close to first-principles accuracy---provided
the model is well trained and uses suitable descriptors---at a fraction of
the cost and with favorable parallel scaling, especially in the weak-scaling
limit.  Related efforts apply the same methodology to coarse-grained
descriptions in order to bridge the time-scale gaps that limit all-atom
empirical models.  A detailed survey of these methods is beyond the scope of
this article; we refer the reader to recent reviews of the state of the
art\cite{mlip_roadmap,mlip_guide}.

\subsection{Create LAMMPS Inputs With AI}
\label{sec:ai-usage:input}

There have been several attempts to build LLM-based assistants that act
like an expert system to help inexperienced users, or users who are not
domain experts, set up simulations with LAMMPS.  This is an area of
active research, but what has been made public so far for use with LAMMPS
is quite limited, and in our experience its usefulness tends to be
reported more optimistically than it warrants.  These assistants work
best for simple, self-contained workflows, but those could often be
equally well assembled by adapting an input deck from a related tutorial.

The biggest challenge for customized LAMMPS AI chatbots is that---due to
the longevity of LAMMPS---the public web is dominated by LAMMPS
tutorials and mailing list or forum discussions that are quite old and
thus potentially outdated\footnote{the LAMMPS mailing list archives go
  back to 2005}.  While a lot of the core functionality of LAMMPS has
been stable for a long time, the refactoring efforts in recent years
described in Sect.~\ref{sec:refactor} can lead to inconsistencies.
Furthermore, the correspondence between the names of internal variables
or functions and the names of the corresponding user-facing keywords is
not always good, and the AI assistants tend to conflate information from
both sources.  The archived inputs themselves are also an unreliable
guide: many of them are \emph{incorrect}, and the corrections appear as
prose or a pointer to the documentation rather than a corrected
input---unlike Stack Overflow, whose answers are complete, reader-ranked
code.

More fundamentally, an LLM has no model of the underlying physics, so it
will readily stitch together fragments from unrelated systems or
incompatible unit systems.  And unlike a syntax error or a crash, such a
flawed physics input is not easily flagged: its results can look entirely
reasonable and reveal their flaws only under careful statistical
analysis.

\subsection{Using AI for Code Review}
\label{sec:ai-usage:review}

As of fall 2025 we apply AI-based code review on GitHub to most
submitted pull requests, alongside our automated testing regime
(Sect.~\ref{sec:test}).  This is in part to reduce the workload on the
developers, but we have also found that supplying project-specific
guidance in a \texttt{.github/copilot-instructions.md} file makes the
automated review markedly more useful: it reliably flags the many small
formal and stylistic changes that we require for consistent behavior and
that are tedious to check by hand.  These reviews frequently also
pre-empt problems that would otherwise surface only later from static
code analysis (Sect.~\ref{sec:test:static}), which---because of its
computational cost---is typically run only after a pull request has been
merged, or no more than once every few days.  This is a clear
improvement over our initial experiments with the same tool, which we
had found underwhelming.

\subsection{Using AI for Debugging}
\label{sec:ai-usage:debug}

The usefulness of an AI coding agent for debugging LAMMPS depends strongly
on how much context can be provided: how precisely the failure is described,
how representative and narrow the example input is, and how quickly it runs.
For a well-documented, easily reproduced issue, a coding agent may
identify the origin quickly and provide a meaningful bugfix.  The interaction
with the developer is in such cases reduced to programming style details
or tweaking the documentation.

On the other hand, when given little information and with a difficult to
reproduce issue, coding agents currently tend to pursue invalid
hypotheses and, in difficult cases, to go in circles without reaching a
viable fix.  But even then the exercise can be worthwhile, because the
agent does not share the developer's preconceptions and may examine
error sources that the developer has already ruled out; on several
occasions an agent that could not fix a bug itself produced a transcript
from which an experienced developer recognized the actual cause.  Where
AI agents are most effective is the large class of bugs that stem from
small oversights that are easily overlooked by a human like the
incorrect use of a function or data structure, swapped indices,
copy-and-paste mistakes, off-by-one errors, or the misuse of a
well-understood algorithm---particularly algorithms from applied
mathematics, with which---in our experience---the coding agents are often
strikingly familiar.

\subsection{Using AI for Code Refactoring}
\label{sec:ai-usage:refactor}

We have found refactoring to be one of the areas where coding agents are
most effective, largely because the externally visible behavior is meant
to stay unchanged while only the implementation is modified---for example to
adopt modern C++ constructs or a different data layout.  A recurring
pattern is that the agent writes a Python script to perform the edits
rather than editing the source directly; for purely mechanical changes
this often works well.  As an illustration, the \texttt{Variable} class
was refactored from several parallel lists into a single container of
per-variable structures with the help of a coding agent.  The bulk of
the mechanical substitution was completed quickly (the coding agent
created and debugged a Python script to do the task), but the result was
only a starting point. The cleanup that followed---reducing redundant code and improving
readability with more modern C++---was done by hand.  This matches our broader experience that
scripted edits are best treated as a first pass to be refined by an
experienced developer who knows how the code has evolved.  The
general-purpose AI used here proved serviceable for such straightforward
tasks but not for writing new or complex code.  Other applications that
were successful are the modernization of internal APIs: some of the
internal data structures in the core C++ classes still somewhat resemble
the data layout of the original LAMMPS implementation in Fortran using flat
arrays.  Replacing those with containers and providing accessor
functions returning pointers to objects instead of the index in the list
is a refactoring task that can often be implemented incrementally and is
well suited to delegate to a coding agent as soon as a sufficiently
large number of adapted cases exist.

\begin{lstlisting}[float=tb,caption={Result of consolidating the
 embedded-atom-method pair styles (Sect.~\ref{sec:ai-usage:refactor}):
 once the shared file readers and force machinery were moved into the
 \texttt{PairEAM} base class, each variant became a constructor-only
 subclass that merely selects its file format and options.  The same
 holds for the accelerated variants, where the duplication---and thus the
 saving---was largest.},label={lst:eam}]
// pair_eam_alloy.h -- the complete class declaration
class PairEAMAlloy : public PairEAM {
 public:
  PairEAMAlloy(class LAMMPS *);
};

// pair_eam_alloy.cpp -- the complete implementation
PairEAMAlloy::PairEAMAlloy(LAMMPS *lmp) : PairEAM(lmp) {
  fileformat = SETFL;   // eam/alloy reads a "setfl" file
  one_coeff = 1;
}

// eam/fs differs only in the file format it selects ...
PairEAMFS::PairEAMFS(LAMMPS *lmp) : PairEAM(lmp) {
  fileformat = FS;
  one_coeff = 1;
}

// ... and eam/he builds on eam/fs, adding a single flag
PairEAMHE::PairEAMHE(LAMMPS *lmp) : PairEAMFS(lmp) {
  he_flag = 1;
}
\end{lstlisting}

A second, more substantial example is the consolidation of the embedded
atom method (EAM) family of pair styles.  Over the years this family had
grown to include several closely related variants (\texttt{eam},
\texttt{eam/alloy}, \texttt{eam/fs}, \texttt{eam/he}, and
\texttt{eam/cd}), each reimplemented in the OPT, OPENMP, INTEL, GPU, and
KOKKOS accelerator packages.  The original design dates back to the early
days of the C++ version of LAMMPS and did not anticipate this later
proliferation of accelerator back ends, so the variants had accumulated a
large amount of nearly identical, duplicated code.  With the help of a
coding agent, the potential-file readers and the special embedding-table
treatment of \texttt{eam/he} were moved into the shared \texttt{PairEAM}
base class, reducing each format-and-engine combination to a small
subclass that only selects the appropriate options in its constructor
(Listing~\ref{lst:eam}).
This removed roughly 5,700 lines of duplicated code without changing any
results, as confirmed by the existing force-style tests for all affected
variants (Sect.~\ref{sec:test:unittest}).  The reductions were largest
for the accelerated variants, which had each duplicated the file readers
in package-specific form; the GPU \texttt{eam/alloy} implementation, for
example, shrank from about 570 to roughly 60 lines.  As in the previous example,
the bulk of the change was the
mechanical relocation of existing, human-written code, while the new
``glue'' code---the file-format dispatch, a shared helper for the
embedding-table lookup, and the constructor-only subclasses---was
generated by the agent under the direction of, and reviewed by, an
experienced developer.\footnote{\url{https://github.com/lammps/lammps/pull/5031}}
This long-overdue cleanup paid off immediately: a follow-up
change\footnote{\url{https://github.com/lammps/lammps/pull/5032}} could
then add accelerated variants of \texttt{eam/he} for the OPENMP, KOKKOS,
GPU, and OPT packages with very little additional effort, again as
constructor-only subclasses built on the now-shared implementation.

Refactoring is not confined to source code: improving and updating
existing \emph{documentation} is another area in which coding agents are
very helpful.  This is especially valuable for scientific software, whose
developers are often more interested in the new science a feature
enables than in writing and maintaining its documentation---with the
result that such documentation has a reputation for being hard to read,
outdated, or incomplete (``use the source\ldots'').

\subsection{Using AI for Writing New Code}
\label{sec:ai-usage:develop}

Writing genuinely new code is, in our assessment, the most demanding
application of AI in LAMMPS, and the one currently advancing most
rapidly.  With frequent model releases, multiple concurrent agent
architectures, and extended ``reasoning'', the most capable models can
produce impressive results---but only if the cost of access to those
frontier models can be met, which can be an obstacle for academic groups with restricted
funding (Sect.~\ref{sec:ai-usage:outlook}).  The decisive
prerequisites for success are a detailed, iterated plan that captures
the algorithmic requirements of a feature, pointers to relevant
publications, pointers to existing source files that can serve as
templates, and---as with code review---a running record of lessons
learned that is made available to the agent in later sessions.

Here are brief discussions of a few representative recent cases.

\begin{enumerate}
\item \label{sec:ai-usage:develop:symmetry} A long-standing feature
  request for enforcing crystal symmetries with an associated
  publication\cite{symd_paper} and a reference
  implementation\cite{symd_code} in C was programmed with a coding agent
  from a plan in a few hours while the agent was mostly working
  unattended.  After a few adjustments to include statements, error
  messages, and documentation, the resulting code closely resembles
  LAMMPS code that was hand-written by an experienced LAMMPS developer.
  The effort to integrate such a contribution is therefore comparable to
  integrating code written by regular (human) contributors to LAMMPS and
  \emph{less} than integrating work by a first time contributor that is
  unfamiliar with the conventions and preferences employed by the core
  LAMMPS developers.

\item \label{sec:ai-usage:develop:ilves} A feature request was submitted
  by researchers that had implemented a constraint solver
  (ILVES\cite{ILVES}) that overcomes the limitation of the SHAKE
  implementation\cite{shake} that only small clusters can be constrained and not
  more than two connected bonds.  In addition to the publication, an
  implementation in another MD code, GROMACS\cite{gromacs}, was made
  available in a public git repository\cite{ILVES_GROMACS_GITHUB}.  This
  was a very demanding task that required multiple revisions and pushed
  the coding agent to its limits.  Since there are significant
  differences in the data layout and parallel architecture between
  LAMMPS and GROMACS, a simple transfer of the code was not possible and
  ultimately two variants of an implementation were obtained: one that
  would function in all use cases, but has large memory consumption
  through using replicated global data and thus bad scaling with system
  size and number of processors and another with distributed memory and
  improved parallel scaling, but restrictions in the support for large
  clusters of constrained bonds, which can be in part alleviated by
  increasing the ghost atom region and thus sacrificing some
  performance.

\item \label{sec:ai-usage:develop:grantest} LAMMPS had very little test
  coverage for discrete element models (DEM).  The coding agent was then
  asked to build a test infrastructure and populate it with a series of
  tests.  Here the coding agent was given the code skeleton and test
  architecture used for molecular and materials modeling systems.  It
  was also pointed to several benchmark publications\cite{gran_bench1,
    gran_bench2, gran_bench3, gran_bench4} and the website of a DEM
  software\cite{mfix_web}, which lists benchmark cases with analytical
  solutions for regression testing and validation.  This turned out to
  be a more straightforward task (apparently the training data for the
  LLM contained significant relevant information), but still was running
  for a long time with many stages because of the size and scope of the
  task.  Again, careful planning, ample reference data, and providing
  specific directions---especially during planning---turned out to be
  crucial for success.  As a bonus, the implementation of the tests
  revealed a bug in the implementation of one of the DEM model
  components that was then subsequently fixed.

\item \label{sec:ai-usage:develop:porting} Porting existing
  functionality to an accelerated package such as KOKKOS or OPENMP is a
  particularly attractive target, since suitable skeleton code can be
  extracted from already-ported classes and the new functionality
  replaces the one from the template.  Asking the coding agent to record
  any lessons learned from successful implementations helps to improve
  the future success rate.  Same as for code review, those instructions
  should contain descriptions of coding style requirements and
  documentation conventions so the need for manual polishing of the
  created code and documentation afterwards is reduced.  With this
  approach the coverage of models supported by the two packages has been
  significantly increased in a short time with rather little effort by
  the LAMMPS developers.  However, it has been crucial that the affected
  models are covered by the testing infrastructure since this kind of
  pattern based adapting of code can lead to subtle bugs when the model
  chosen as a template for a newly ported model has small but
  significant differences, e.g. in how energy or stress contributions
  are tallied.
\end{enumerate}

The broader lesson is that a coding agent, steered by an experienced
LAMMPS programmer, can now produce a near-complete implementation of a
requested new feature, provided the researcher can supply sufficient
context---one or more publications, reference code in another code base,
or pointers to similar implementations in the LAMMPS code base---and the
task is of low to moderate complexity.  Conversely, careless or uncritical
use of coding agents---or not using the most advanced models---can lead to
poor results, such as source code that is hard to read and understand, or
an implementation that is incorrect in ways that are not obvious or easy
to detect.

\subsection{Using AI for Writing Documentation and Tutorials}
\label{sec:ai-usage:docs}

Because large language models are first and foremost language models,
writing documentation is a particularly natural application, and one
that complements the improvement of existing documentation noted above
(Sect.~\ref{sec:ai-usage:refactor}).  The LAMMPS manual is large and
maintained as reStructuredText processed with Sphinx
(Sect.~\ref{sec:docs}), and keeping it complete, consistent, and
readable is a continuous effort.  AI assistants can draft a first
version of a reference page for a new command or style from the source
code, its comments, and the integrated Python docstrings and Doxygen
markup (Sect.~\ref{sec:docs:doxygen}); they can adapt an existing page
for a closely related feature; and they are effective at polishing
language and enforcing a consistent style, which is especially helpful
because many contributions come from authors who are not native English
speakers.

Tutorials and longer worked examples are a second area where such
assistance is useful.  Here a model can propose a structure, turn terse
notes into connected prose, and add explanatory text around an existing
input deck.  As with the construction of input scripts
(Sect.~\ref{sec:ai-usage:input}), however, the scientific content---the
choice of model, the physical reasoning, and the recommended best
practices---must be supplied and verified by a domain expert; the
language model contributes primarily the presentation, not so much the
science.

The central caveat is that documentation is only useful when it is
correct, and a plausible-sounding but wrong description can be more
damaging than an obvious omission because users rely on the manual.
Language models readily invent command keywords, default values, or
units that do not exist, or describe behavior that does not match the
actual code, so every AI-generated or AI-edited page has to be checked
against the implementation by someone familiar with it.  Some classes of
error are caught automatically by the documentation-consistency scripts
described in Sect.~\ref{sec:docs}, which verify, for example, that every
command and style is documented, appears in the index and the command
summary tables, and that cross-references resolve---but these checks
detect only formal omissions, not statements that are merely untrue.  As
elsewhere in this section, the productivity gain therefore comes with a
shift of effort toward careful review (Sect.~\ref{sec:ai-usage:drawbacks}).

\subsection{Risks and Drawbacks}
\label{sec:ai-usage:drawbacks}

The rapid adoption of AI coding tools has been accompanied by a growing
awareness of their risks.  A widely voiced concern, which matches our own
experience, is that large language models can produce plausible-looking
code that is subtly incorrect or insecure, that uncritical reliance on them can erode the skills and
understanding of developers, and that their output inevitably reflects the
errors and biases present in their training data.  For scientific software
these concerns are amplified: correctness is paramount yet often hard to
verify, the results feed directly into published research, and a mistake
that merely looks reasonable can invalidate scientific conclusions without
ever triggering a crash or a failed test.  In our view, AI coding tools are
therefore most safely used as an accelerator for experienced developers who
can critically evaluate the output, not as a substitute for that expertise.

A more immediate drawback is cost.  Productive use of coding agents requires
access to the most capable ``frontier'' models, which are available only
through paid subscriptions or metered API access.  For an individual
academic researcher---typically already struggling to secure funding---a
subscription large enough to sustain full-day use is a non-trivial expense
that is hard to justify on a grant budget and, because access is usually
accounted per seat, hard to share across a group.  The problem is more acute
for open-source projects such as LAMMPS that are maintained largely by
volunteers and have no dedicated budget for such tools.  There is a real
risk that the productivity gains from AI accrue preferentially to
well-funded groups and widen the gap to community-driven projects.

A second, and so far unresolved, question concerns authorship and licensing.
A coding agent produces its output by drawing on patterns learned from a
vast body of source code, including many open-source projects whose licenses
impose differing and sometimes incompatible terms.  It is therefore not
always clear whether a given piece of AI-generated program text is an
original creation or, in effect, a copy of existing code whose license ought
to be honored.  This matters in particular for a project like LAMMPS, which
is distributed under a specific license (GPLv2) and must be able to vouch for
the provenance of the code it ships.  Lacking clearer legal and ethical
guidance, we currently subject AI-assisted contributions to the same
provenance and licensing scrutiny as any other contribution.

Finally, there is the matter of code quality, sometimes called ``AI slop'':
generated code can be needlessly complex, verbose, or hard to follow, so
that reviewing and reworking it costs more effort than it saves.  Our
observation is that this correlates strongly with how much relevant context
the agent was given to make meaningful decisions, and with how closely the
task resembles other, already-solved problems represented in the training
data.  Even when the input is good and the result is sound, the higher rate
at which code can now be produced shifts the bottleneck rather than removing
it: the burden on the core maintainers moves from writing new code and
fixing bugs toward reviewing contributed code.  For a project like LAMMPS,
with a comparatively small group of core maintainers
(Sect.~\ref{sec:history:expand}), this added review load is a substantial and
growing burden.

\subsection{Outlook}
\label{sec:ai-usage:outlook}

Of the topics discussed in this section, the creation of machine-learning
interatomic potentials is currently the best understood and most common
use of AI in relation to LAMMPS and several mature packages exist even
though there is also still rapid development and more capable, reliable,
and accurate models should be expected.  In comparison, using AI coding
agents for programming and debugging tasks has become viable more
recently and the rate of improvement has been very high.  However, using AI
chatbots to assist users to perform simulations with LAMMPS without
possessing sufficient domain knowledge is still very problematic for
multiple reasons.  This is not unexpected, since for most scientists and
developers the main objective is to advance their research and the
capabilities of the software and not to train their peers (and sometimes
potential future competitors).

The principal barrier to using coding agents productively is to have
access to the most advanced frontier models and sufficient funding for
continued access.  Since coding agents can work largely unattended,
an experienced developer can work with multiple sessions concurrently.

We see the most immediate value in completing the backlog of dormant
projects---feature requests, bug reports, and partial implementations
that have accumulated over the years---where the scientific goal is
clear and a reference or publication is available, but also in general
refactoring and modernization of the code base.  Given the size of the
LAMMPS code and its age, there are still many parts that are
mostly unchanged from the early times and could be implemented
differently in modern C++.  In our experience, coding agents are well suited to such repetitive
tasks, since they apply consistent care across many near-identical cases
where human attention tends to wane.

A promising division of labor is for an experienced developer to work
with the agent in its planning mode to produce an implementation plan,
which can then be reviewed and adjusted.  We also expect to keep
recording lessons learned and refining the project-specific instructions
on file that, as in code review, can make a substantial difference to
the quality of the results.

\section{LAMMPS-GUI}
\label{sec:gui}

\begin{figure}[tb]
    \centering
    \includegraphics[width=\linewidth]{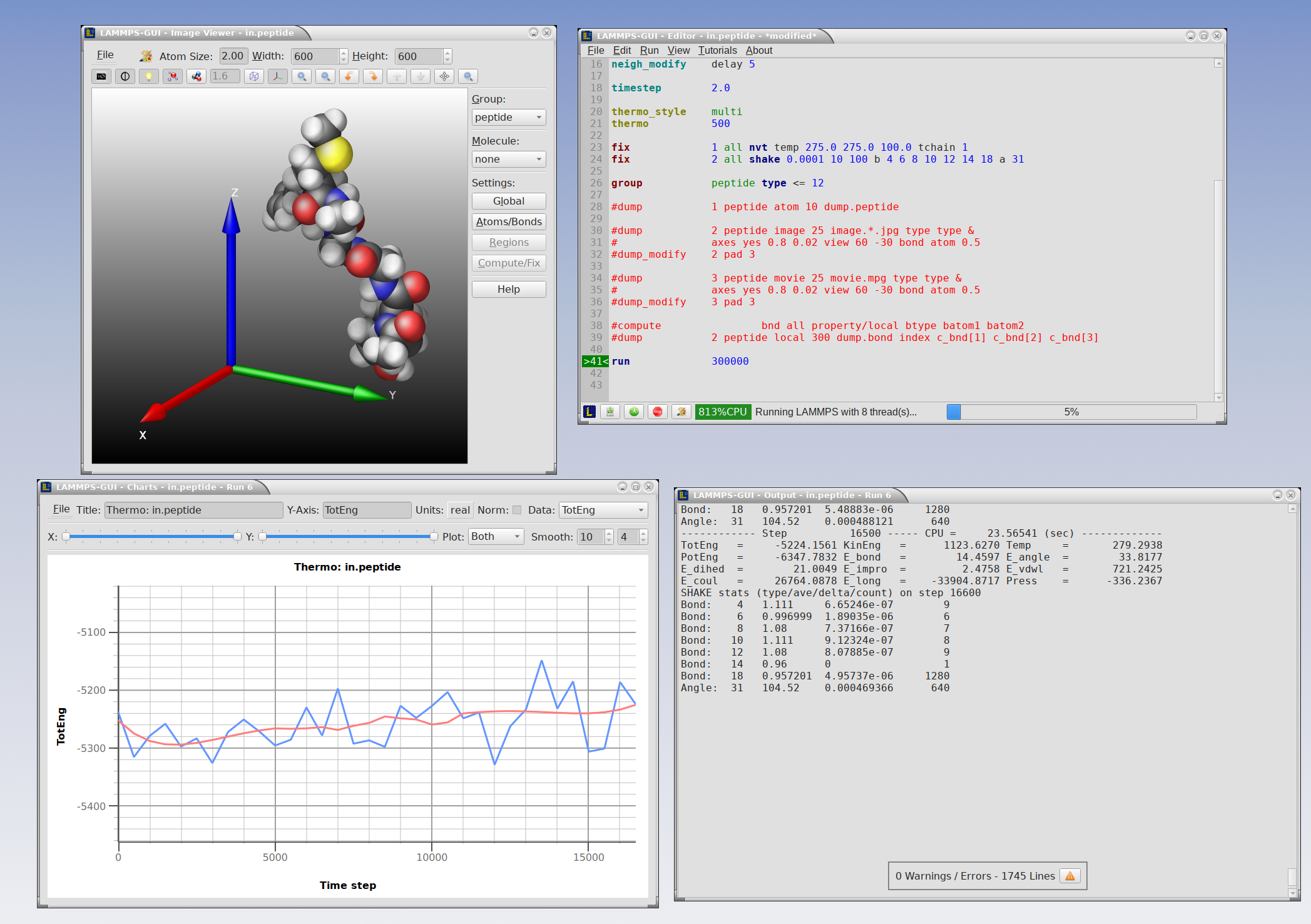}
    \caption{Screenshot of LAMMPS-GUI while running a simulation}
    \label{fig:lammps-gui}
\end{figure}

For many years, LAMMPS would advertise on its homepage that it was a
software deliberately designed without a graphical user interface.  This
proved to be particularly challenging when performing tutorials, since a
significant amount of time had to be spent teaching the participants how
to compile and install LAMMPS and the necessary tools for editing LAMMPS
inputs, extracting results, or visualizing simulated systems, and
teaching aspiring LAMMPS users how to use those tools.  This is further
complicated, as popular operating systems like Windows or macOS ship
without compilers and different, OS-specific tools for editing inputs,
plotting graphs, or visualization have to be introduced.

For some time, this could be worked around by using virtual machine
images of a Linux system for Intel processors with everything configured
and installed, so that only one set of tools would need to be introduced,
and they were guaranteed to be identical for all users.  For several
years the LAMMPS developers provided images for VirtualBox\cite{virtualbox}
and the experience was very positive: the VirtualBox software is available
for all major platforms (Windows, macOS, Linux) and 64-bit x86 CPUs were
the common platform.  To use Linux client virtual machines and thus force
users to learn a few Linux basics in the process is a justified effort,
since the majority of HPC clusters and supercomputers are running Linux,
and thus users will need to learn using Linux and those tools eventually.
However, this approach broke down when Apple switched from Intel CPUs
to their own, ARM-based CPU architecture.

Thus, the approach for focusing on teaching LAMMPS during tutorials was
to write a portable tool that has a graphical user interface and
includes facilities for editing LAMMPS inputs, running LAMMPS,
monitoring the simulation, collecting the thermodynamic output produced,
creating plots from it, and creating visualizations of the system
geometry.  This led to the development of
LAMMPS-GUI\cite{lammps_gui_home}, for which pre-compiled binary packages
for Linux x86\_64, Windows, and macOS are provided for download
\cite{lammps_releases}.  Fig.~\ref{fig:lammps-gui} shows a screenshot of
a simulation running in LAMMPS-GUI, which is also featured in the online
LAMMPS tutorials\cite{lammps_tutorial}.

Although GUI front-ends for LAMMPS existed before---mostly as part of
commercial simulation environments supporting a variety of simulation
tools---these typically run LAMMPS as a separate, external process and can
interact with it only by monitoring and parsing the output files that LAMMPS
writes.  That approach is fragile, since the content and format of LAMMPS
output can change between versions.  LAMMPS-GUI instead integrates much more
tightly: it runs LAMMPS as a separate thread \emph{within the same process}
through the C-language library interface, and uses the introspection
functions of that interface to read the internal data of the running
simulation directly.\footnote{The introspection functions acquire internal
locks, so LAMMPS-GUI waits while LAMMPS is updating its data and never reads
an inconsistent state.}  It therefore does not need to generate or parse
output files in order to monitor a simulation's progress or to capture its
results.

The design of LAMMPS-GUI benefits from the recent refactoring work in
LAMMPS that introduced introspection facilities
(Sect.~\ref{sec:library:c}) and had errors throw exceptions instead of
aborting the process (Sect.~\ref{sec:refactor:exceptions}).  As already
noted there, this is what keeps the GUI from crashing when the LAMMPS
thread aborts on an error; beyond reporting the error, it also indicates
at which step of the input it happened, and the user can then correct the
input so that LAMMPS-GUI re-initializes the LAMMPS object and re-launches
the simulation.

\begin{figure*}[tb]
\centering
\begin{minipage}[b]{0.34\textwidth}\centering
  \includegraphics[width=\linewidth]{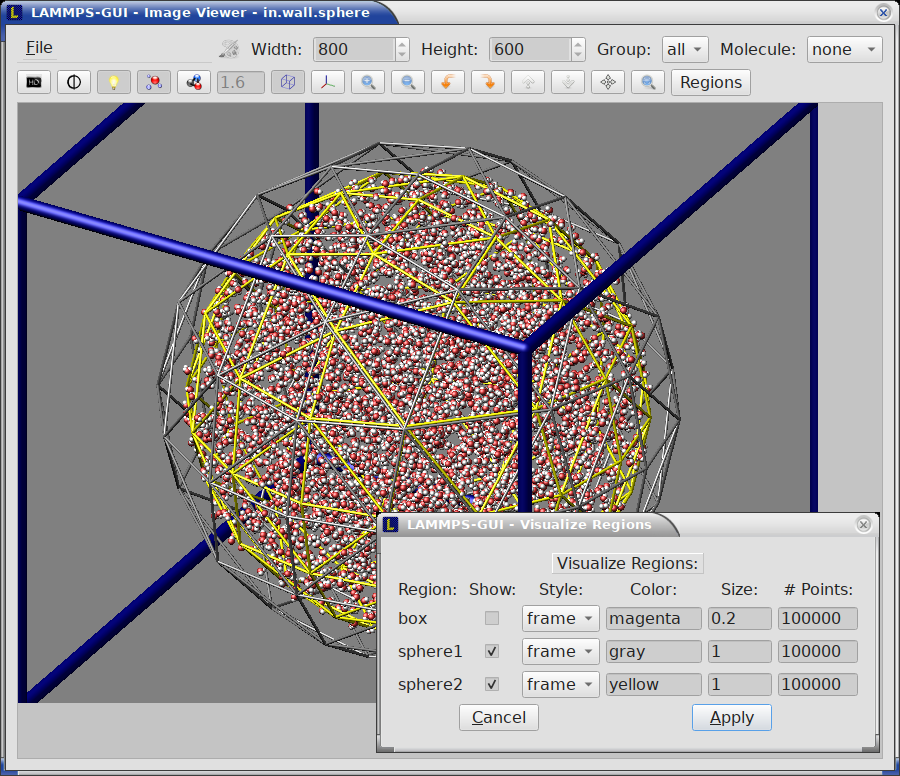}
\end{minipage}\hfill
\begin{minipage}[b]{0.31\textwidth}\centering
  \includegraphics[width=\linewidth]{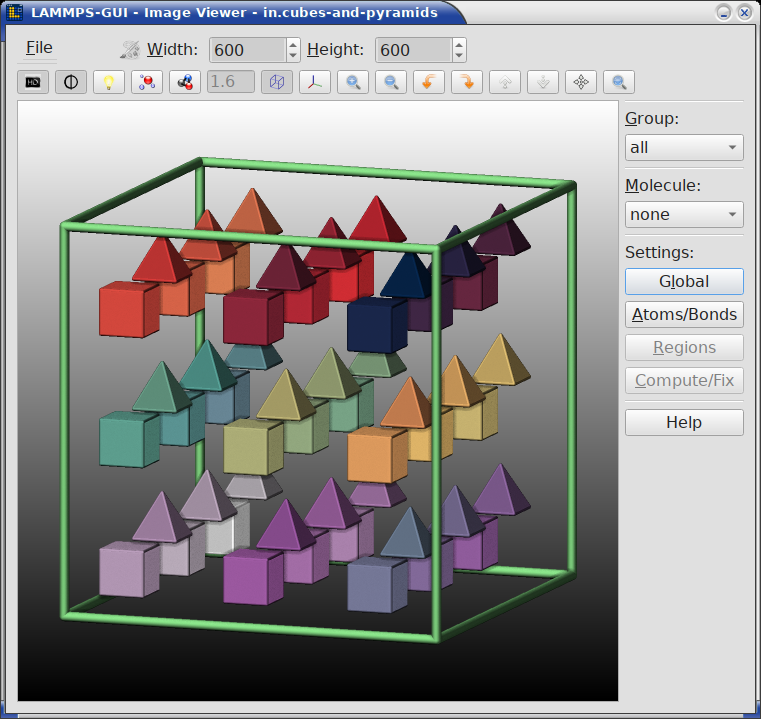}
\end{minipage}\hfill
\begin{minipage}[b]{0.31\textwidth}\centering
  \includegraphics[width=\linewidth]{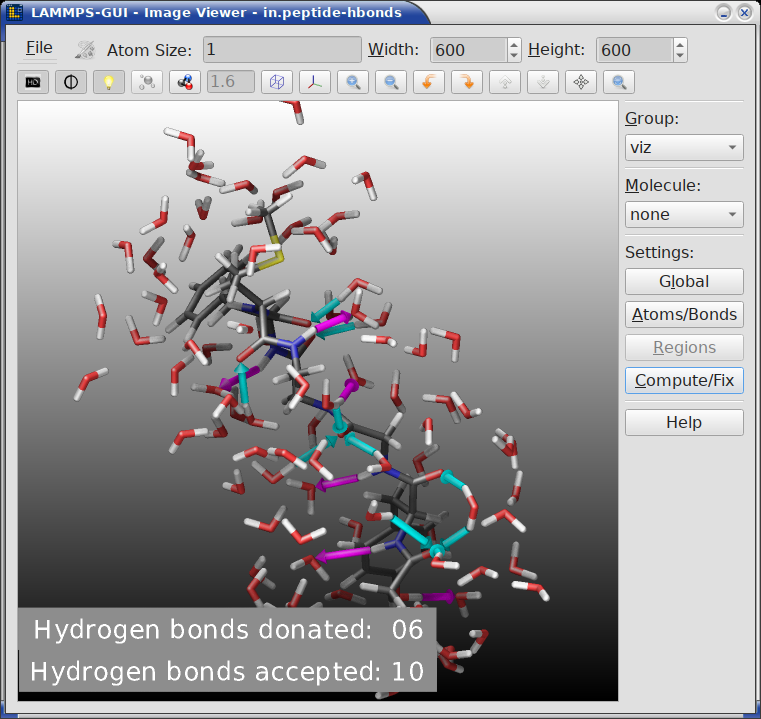}
\end{minipage}
\caption{Visualizations created with LAMMPS-GUI and the new GRAPHICS package.
  Left: two spherical regions shown together with the dialog used to select them
  and adjust the rendering options. Middle: body particles rendered as cubes and
  pyramids. Right: a peptide fragment with its nearest water molecules, arrows
  representing hydrogen bonds, and overlaid annotations.}
\label{fig:gui-viz}
\end{figure*}

Also, implementing LAMMPS-GUI has inspired improvements in the LAMMPS
code like caching of thermodynamic output for advanced dump styles that
store global data in addition to per-atom data, expansion of
multi-threading support (since using the GUI is incompatible with the
default MPI parallelization), and a significant expansion of the
graphics capabilities in LAMMPS culminating in the creation of a new
GRAPHICS package that brings many visualization features to LAMMPS that
are also available in dedicated visualization programs like
OVITO\cite{ovito_home} or VMD\cite{vmd_home} and the writing of a
detailed visualization Howto document\cite{lammps_visualization}.  Some
highlights of the additions are: support for anti-aliasing,
\emph{screen-door} transparency, vertical color gradient background,
rendering text, importing images, visualizing regions, and enhanced
visualization of ellipsoids, superellipsoids, and polyhedral particles.
Figure~\ref{fig:gui-viz} shows three such visualizations.

The enhancements in LAMMPS in turn motivated expanding the visualization
support in LAMMPS-GUI to enable the added graphics features
interactively from custom dialogs and generally provide more detailed
control over the available settings for the dump image command. Most
notable (and most asked for by users) is the ability to enable coloring
of atoms by property and to select between pre-defined color maps.
Since the internally assembled command line for the \texttt{dump image}
and \texttt{dump\_modify} commands used by the LAMMPS-GUI for
creating visualizations is recorded and can be copied into the edited
input file, this provides a convenient way to create a command line
template for a visualization to create images during a run that is
independent from LAMMPS-GUI (e.g. from a parallel run on an HPC
cluster).  Building such a command line that has correct syntax is quite
challenging since the \texttt{dump image} command has many options and
keywords and many are quite complex.

\section{Summary}
\label{sec:summary}

In this paper, we analyze how various modern software development
methods and tools help to improve the LAMMPS molecular dynamics software
package and its development process.  It would not be possible today to
develop and maintain LAMMPS with the same efficiency and the same level
of code quality, had we continued with the historical development
process.

Moving the source-code management to GitHub made the process more
transparent and redistributed effort from the developers who integrate
contributions to the contributors themselves, supported by explicit
policies and conventions that spread the remaining maintenance and
integration work across the core team.  Automated testing and static
code analysis are central to this balance: most checks now run
automatically---and therefore more consistently than when performed by
hand---and give contributors immediate feedback on each pull request, so
that programming errors are flagged during development rather than
discovered much later by users.  Static analysis was noisy at first,
producing many false positives because of the design and programming
conventions of LAMMPS, but the rate of new reports has since become
small and easy to triage.  One visible effect is a shift in the kind of
problems that remain: bugs reported in stable releases are now mostly in
recently added features, while long-established code is rarely
implicated, and a growing fraction of reports stem from
misunderstandings or errors in the documentation.  Overall, this
approach strikes a good balance between integrating new features quickly
and maintaining correctness, consistency, and code quality without
placing an excessive review and maintenance burden on the core
developers.  These same practices---review before merge, automated
testing, and static analysis---are also what allow the project to address
software security and the integrity of its supply chain
(Sect.~\ref{sec:security}), from catching memory-safety defects before a
release to vouching for the provenance of the sources and the released
binaries.

Many of the improvements are ``under the hood'', that is, in the
implementation of the functionality without changing the interface
(much) or in the development process itself so that developers can
better cope with the growing popularity and the correspondingly growing
volume of contributions to LAMMPS. Other changes are aimed at the
perceived change in the kind of users from people with experience in
programming to people who lack that experience and use pre-compiled
executables.  With the improvement to error messages and the addition of
LAMMPS-GUI specifically, the needs of beginner- and intermediate-level
users are addressed. The new ``Programmer's Guide'' targets developers
instead and tries to improve their understanding of development
strategies and provide documentation of added utilities and portability
abstractions.  Many of these changes to the source code, its library
interfaces, and its documentation are interconnected and benefit from
each other.

Underlying all of these efforts is a more general lesson: to remain
relevant over decades, a software package must continually adapt to
changes in hardware and software and adopt best practices as they
evolve.  This works best with a sufficiently large team of engaged
developers with diverse expertise across the relevant research areas and
software engineering, who are nonetheless able to find compromises that
serve the majority of users and developers.  The continued growth in the
popularity of LAMMPS---evident in its citation count
(Fig.~\ref{fig:citations}) and in the growth of the code base
(Figs.~\ref{fig:git_lines} and \ref{fig:git_commits})---suggests that
the core development team has chosen a good path to remain relevant in
the future.

\section{Future Plans}
\label{sec:future}

Moving forward, the current plan is to continue the refactoring process
(see Sect.~\ref{sec:refactor}) and gradually replace old-fashioned and
error-prone code constructs with simpler and modern ones, as long as the
changes do not negatively affect the serial and parallel performance of
LAMMPS or the readability of the code.  This will significantly benefit
from the availability of capable AI coding agents (see
Sect.~\ref{sec:ai-usage:refactor}) and thus we expect more progress in
this area.  When carefully steered by an experienced developer, and
given suitable tasks, current generation coding agents can work largely
unattended and produce refactored code similar to what an experienced
developer would write.

In that process also the minimum C++ standard version and CMake version
requirement will have to be regularly reconsidered on the basis of which
versions are still widely in use, specifically on major supercomputers.
Since we advanced from requiring C++11 in 2020 directly to requiring
C++17 in 2025, there should be a longer period until we can require
C++20.  As guideline we look at what is available in the oldest
supported Ubuntu LTS and Red Hat Enterprise Linux versions.  The legacy
build system has already been reduced to packages without complex
maintenance requirements, so a full removal is not too far away, and
then the CMake system can be made more modular and better aligned with
best practices and common conventions for CMake.

The expansion of the library of tests and the ``Programmer's Guide''
part of the LAMMPS manual are ongoing projects with high priority.  Both
will benefit from a further expansion of the core LAMMPS team as well as
using AI coding agents (see Sect.~\ref{sec:ai-usage:develop}) and thus
an increase of the available workforce.  The challenge for expanding the
test library is that most of the low hanging fruit has already been
collected and thus, even with AI support, increasing the test coverage
requires creativity and significant effort.

We need to spend more effort training aspiring LAMMPS developers so that
contributions are of high quality and require less work integrating but
also so they may eventually be asked to join the core team and take
over maintenance responsibilities and handle project management tasks.
We have to prepare the LAMMPS project for an inevitable transition of
leadership as senior developers are already retired (though still
contributing) or close to retirement age.

A related technical priority is to simplify and make more robust the
mechanism by which external packages are integrated into LAMMPS.  This
is becoming particularly important for machine-learning interatomic
potentials, which are frequently developed as independent libraries that
are interfaced with several molecular dynamics engines and maintained on
their own release schedules.  The current integration requirements are
comparatively fragile, and we plan to provide a less intrusive and more
durable interface so that such external packages can be kept up to date
with less effort for both their developers and the LAMMPS maintainers.

Finally, ensuring the long-term viability of LAMMPS requires making its
maintenance less dependent on computer hardware resources provided by
the institutions that hosting LAMMPS developers (at present Sandia
National Laboratories and Temple University).  This is already achieved
in part by using cloud resources that are provided free of charge to
open-source projects, for example by GitHub.  To place the project on a
more durable footing, we plan to join the High Performance Software
Foundation (HPSF)\footnote{\url{https://hpsf.io/}}, part of the Linux
Foundation, and to make use of the infrastructure it provides---for
example, for project web hosting.

\section*{Data availability}

LAMMPS\cite{lammps_home,lammps_doi} and LAMMPS-GUI\cite{lammps_gui_home}
are open source software and their source code and packages with
pre-compiled binaries are freely available for download on their
respective GitHub repositories and are distributed under the terms of
the GNU Public License Version 2 (GPLv2).

\section*{Acknowledgments}

The author thanks Richard Berger for his code contributions, guidance,
and advice while working on modernizing LAMMPS and its development
process.  Thanks also go to Steve Plimpton and Aidan Thompson for
providing detailed feedback and encouragement during the writing of this
article and for first-hand information on the early days of LAMMPS
development.

\textbf{Funding:} Financial support provided by Sandia National
Laboratories under POs~2149742 and 2407526 is gratefully acknowledged.

\balance

\bibliographystyle{elsarticle-num}
\bibliography{modernizing-lammps_long}
\vfill
\end{document}